%% file: main.tex
\documentclass[12pt, a4paper, oneside]{book}

\usepackage{amsmath, hyperref, setspace, graphicx, caption, subcaption}
\usepackage[a4paper, top=1in, bottom=1in, left=1in, right=1in]{geometry}
\graphicspath{ {./02_Images/}}
\usepackage[export]{adjustbox}
\usepackage[style=apa, sorting=nyvt]{biblatex}
\hypersetup{
  colorlinks = true,
  citecolor = black,
  urlcolor = blue,
  linkcolor=black,
  pdfauthor = {Kenil Ajudiya},
  pdftitle = {Interpreting Galaxy Physical Properties Using Stellar Population Synthesis},
  pdfpagemode = FullScreen,
  bookmarksopen = true}

\newcommand{\partref}[1]{Part \ref{#1}}
\newcommand{\chref}[1]{Chapter \ref{#1}}
\newcommand{\secref}[1]{Section \ref{#1}}
\newcommand{\figref}[1]{Figure \ref{#1}}
\newcommand{\rnum}[1]{\uppercase\expandafter{\romannumeral #1\relax}}
\newcommand{\blankpage}{
    \begin{center}
        \vspace*{150pt}
        \textit{This page intentionally left blank}
        \thispagestyle{empty} \addtocounter{page}{-1} \newpage
    \end{center}}

\begin{document}
    
    \singlespacing
    \input{01_Chapters/01_titlepage}
    
    \blankpage
    
    \begin{center}
        \vspace*{150pt}
        \thispagestyle{empty}
        Dedicated to my parents,\\
        Mr. and Mrs. Ajudiya
    \end{center}
    \newpage
    
    \blankpage
    
    \frontmatter
    \onehalfspacing
    
    \input{01_Chapters/02_acknowledgements}
    \newpage
    
    \input{01_Chapters/03_declaration_and_certificate}
    \newpage
    
    \blankpage
    
    \input{01_Chapters/04_abstract}
    \newpage
    
    {\hypersetup{linkcolor=black} \tableofcontents}
    
    {\hypersetup{linkcolor=black} \listoffigures}
    
    
    \mainmatter

    \input{01_Chapters/ch01_intro}

    \part{Stellar Population Synthesis}\label{part1}
        \input{01_Chapters/ch02_SSP}
        \input{01_Chapters/ch03_CSP}
        
    \part{Inferring Galaxy Physical Properties}\label{part2}
        \input{01_Chapters/ch04_MAGPHYS}
        \input{01_Chapters/ch05_conclusion}
    
    \printbibliography
    \addcontentsline{toc}{chapter}{Bibliography}
    
\end{document}

%% file: 01_Chapters/01_titlepage.tex
\begin{titlepage}
    \begin{center}
        
        \Huge{Interpreting Galaxy Physical Properties\\
        Using\\
        Stellar Population Synthesis}
        
        \vspace{30pt}
        
        \large{\textit{Submitted in partial fulfillment of\\
        the requirements for the award of the degree of}}
        
        \vspace{15pt}
        
        \LARGE{Bachelor of Science (Research)\\
        in\\
        Physics}
        
        \vspace{30pt}
        
        \normalsize{by}
        
        \vspace{10pt}
        
        \large{Kenil Ajudiya}\\
        \normalsize{S.R. No. 11-01-00-10-91-20-1-18711\\
        Undergraduate Programme\\
        Indian Institute of Science}
        
        \vspace{10pt}        
        
        \begin{figure}[!h]%
            \centering
            \subfloat{{\includegraphics[height=5cm]{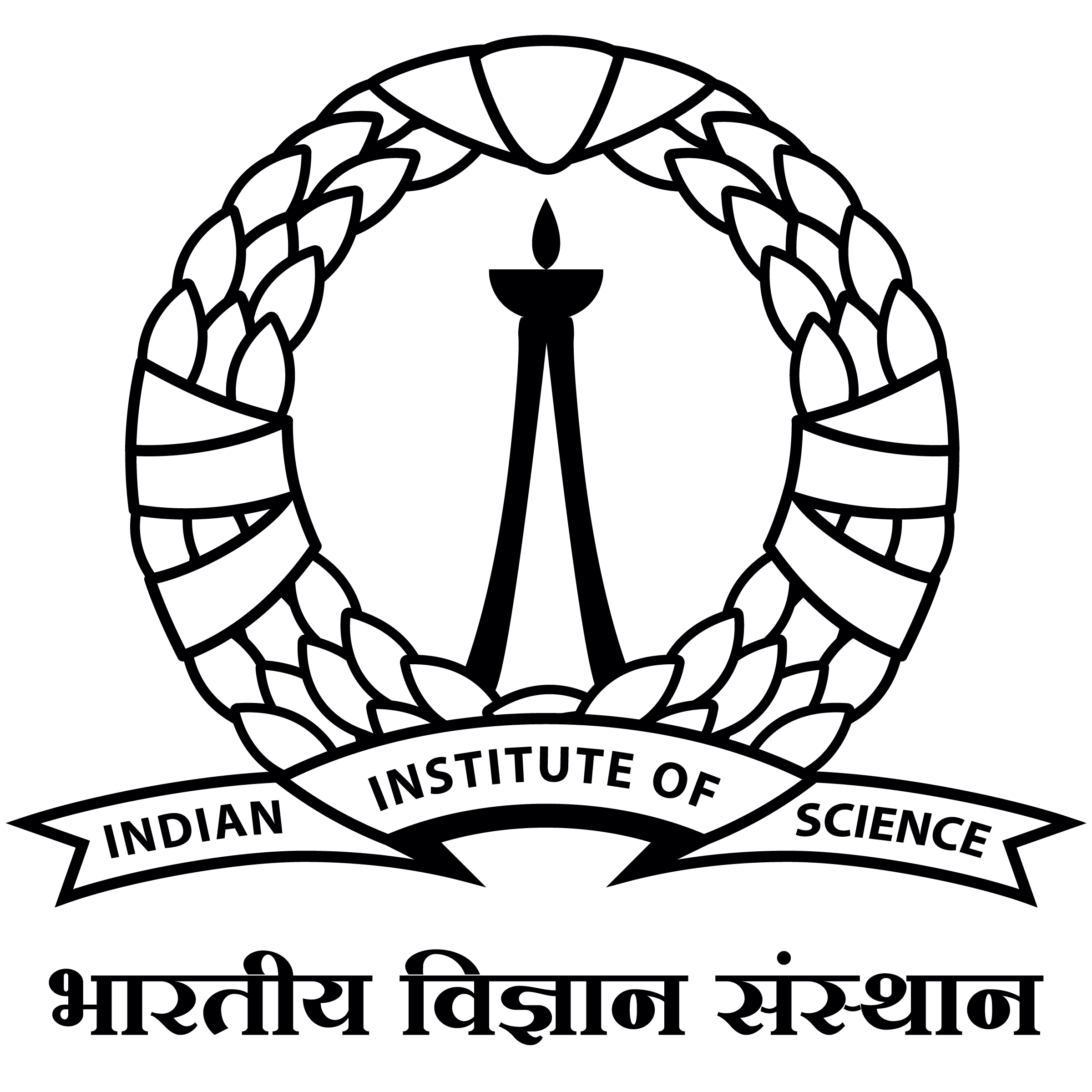} }}%
            \qquad
            \subfloat{{\includegraphics[height=5cm]{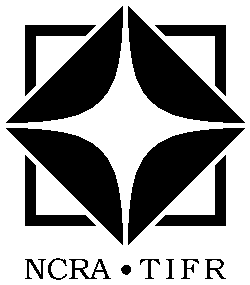} }}%
        \end{figure}

        \vfill

        \normalsize{Under the supervision of}
        
        \vspace{10pt}
        
        \large{Prof. Yogesh Wadadekar}\\
        \normalsize{National Centre for Radio Astrophysics - \\
        Tata Institute of Fundamental Research (NCRA-TIFR), Pune.} 
        
        \vspace{5pt}
        \normalsize{and}
        \vspace{5pt}
        
        \large{Prof. Nirupam Roy}\\
        \normalsize{Department of Physics,\\
        Indian Institute of Science (IISc), Bengaluru.}
        
        \vspace{10pt}
        
        \normalsize{April 15, 2024}
        
    \end{center}
\end{titlepage}

%% file: 01_Chapters/02_acknowledgements.tex
\chapter*{\centering Acknowledgements}
    \addcontentsline{toc}{chapter}{Acknowledgements}
    Life, at IISc, has changed in countless ways, fortunately, for the better. The role of many people in this change makes me feel grateful and blessed. First and foremost, I would like to thank my mentors Prof. Nirupam Roy, Prof. Yogesh Wadadekar, Prof. Aloke Kumar, Prof. Chetan Singh Thakur, Prof. Visweshwar Ram Marthi, Dr. Prabu Thiagaraj and Prof. Chandni Usha, all of whom I was fortunate to have worked with on fruitful research endeavours. Lectures on particle physics by Prof. Nirmal Raj, those on the interstellar medium by Prof. Biman Nath, those on quantum mechanics by Prof. Baladitya Suri and those on ecology and animal behaviour by Prof. Rohini Balakrishnan are now memories for me to cherish for the lifetime! Exciting discussions with Manish Tamta and Avinash Paladi on micro-lensing by primordial black holes and X-ray pulsars will continue till eternity. I am thankful to both of them and many other doctoral and graduate students at IISc and at NCRA who have broadened the horizons of my knowledge and understanding through numerous discussions and sometimes, through debates as well. I acknowledge the contributions by Pralay Biswas towards my Bachelor's thesis project, especially the comments and suggestions he gave while writing of this thesis. I am fortunate to have friends like Fida Fathima, Hemansh Shah, Arzoo Kathewadi, Kruti Bhingradiya, Satyapreet Singh Yadav, Sahil Nandi, Pranav Kalsi, Dibyashanu Pati, Balkrishna Sharma and many others, who were there to share the joy in my highs and as a supportive pillar for my mental well-being in my lows. I would like to acknowledge the financial support I received from the Department of science and Technology (DST), Government of India (GOI), via the Kishore Vaigyanik Protsahan Yojna (KVPY).

    \noindent Lastly, I would like to thank two of the most important people in my life, my mother and my father. None of this would have been possible without their constant support and belief in my pursuits. Thank you for having faith in me and for being there for me as I went through the ups and the downs of my life.
    
    \vspace{15pt}
    \noindent Kenil Ajudiya\\
    April 15, 2024\\
    NCRA-TIFR, Pune. 411007

%% file: 01_Chapters/03_declaration_and_certificate.tex
\chapter*{\centering Declaration}
    \addcontentsline{toc}{chapter}{Declaration}
    
    \noindent I, Kenil Ajudiya (SR. No. 1-01-00-10-91-20-1-18711), hereby declare that this thesis entitled "Interpreting Galaxy Physical Properties Using Stellar Population Synthesis", submitted in partial fulfillment of the requirements for the award of the degree of Bachelor of Science (Research) in Physics, is a presentation of my original research work done under the guidance of Prof. Yogesh Wadadekar at the National Centre for Radio Astrophysics - Tata Institute of Fundamental Research (NCRA-TIFR), Pune, and Prof. Nirupam Roy of the Department of Physics at the Indian Institute of Science (IISc), Bengaluru, and that it has not been submitted elsewhere for any degree or diploma. Wherever contributions of others are involved, every effort is made to indicate this clearly, with due reference to the literature, and acknowledgement of collaborative research and discussions.
    
    \begin{flushright}
        \includegraphics[scale=0.2]{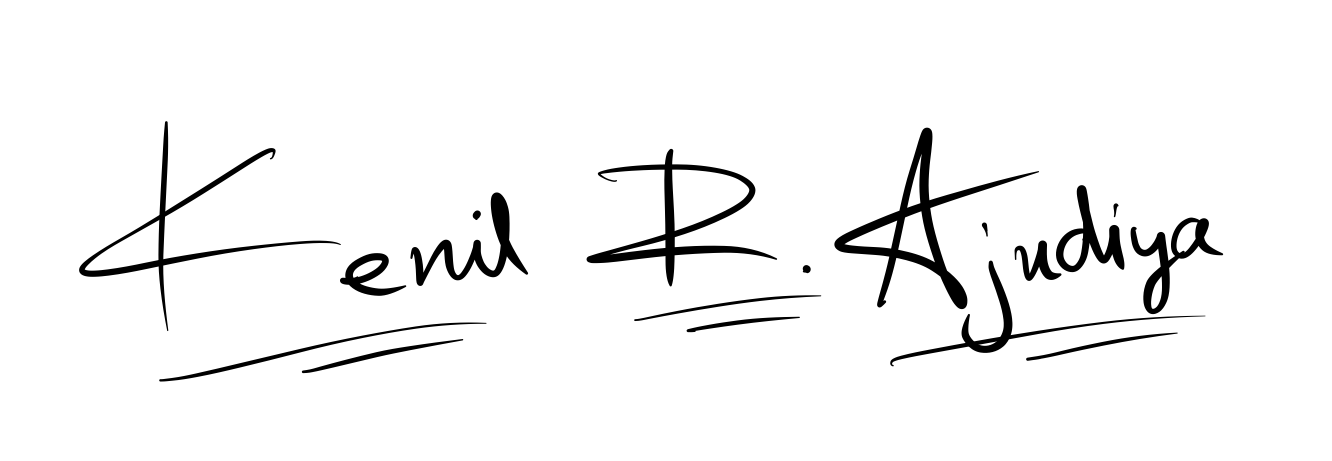}\\
        (Kenil Ajudiya)\hbox{\hskip 25pt}\\
    \end{flushright}
    
    \noindent Place: NCRA-TIFR, Pune. 411007\\
    \noindent Date: April 15, 2024

\chapter*{\centering Certificate}
    \addcontentsline{toc}{chapter}{Certificate}
    
    \noindent This is to certify that the work contained in this thesis entitled "Interpreting Galaxy Physical Properties Using Stellar Population Synthesis", submitted by Kenil Ajudiya in partial fulfillment of the requirements for the award of the degree of Bachelor of Science (Research) in Physics, has been carried out by him under my supervision and under the nominal supervision of Prof. Nirupam Roy, in part at the National Centre for Radio Astrophysics - Tata Institute of Fundamental Research (NCRA-TIFR), Pune, and at the Department of Physics at the Indian Institute of Science (IISc), Bengaluru, and no part of it has been previously submitted for a degree, diploma or any other qualification at this university or any other institution to the best of my knowledge.
    
    \vspace{1in}
    \hspace{25pt} (Prof. Yogesh Wadadekar)
    
    \vfill
    
    \noindent This is to certify that the work contained in this thesis entitled "Interpreting Galaxy Physical Properties Using Stellar Population Synthesis", submitted by Kenil Ajudiya in partial fulfillment of the requirements for the award of the degree of Bachelor of Science (Research) in Physics, has been carried out by him under the supervision of Prof. Yogesh Wadadekar and under my nominal supervision, in part at the National Centre for Radio Astrophysics - Tata Institute of Fundamental Research (NCRA-TIFR), Pune, and at the Department of Physics at the Indian Institute of Science (IISc), Bengaluru, and no part of it has been previously submitted for a degree, diploma or any other qualification at this university or any other institution to the best of my knowledge.

    \vspace{1in}
    \hspace{25pt} (Prof. Nirupam Roy)\hbox{\hskip 25pt}

    \vspace{10pt}
    \noindent Date: April 15, 2024

%% file: 01_Chapters/04_abstract.tex
\chapter*{\centering Abstract}
    \addcontentsline{toc}{chapter}{Abstract}
    Galaxy formation and evolution is one of the most active areas of research in astronomy. In recent times there have been several developments on the observational fronts particularly with the discovery of several relations between galaxy physical properties. The exact details of how they come about still remains to be understood. Such a development has been primarily possible due to a deluge of multi-wavelength data ranging from the ultra-violet (UV) to the radio, mainly due to wide field surveys e.g., the Sloan Digital Sky Survey (SDSS) in the optical. Meanwhile, simultaneous theoretical developments like a better understanding of dust attenuation and emission led to the development of techniques to extract information from the SEDs of galaxies, exploiting information from the far-ultraviolet (FUV) to the far-infrared (FIR). The substantial progress made in stellar evolution theory in the 1980s and 1990s paved the way for the latter approach to become the de facto standard in modeling the SEDs of galaxies. It became possible to synthesise a population of stars with a certain distribution and evolve it in time, keeping track of the emission from the stars, new star formation activity, gas enrichment with elements heavier than hydrogen and helium, and the absorption and re-emission from the interstellar dust. This technique, known as the stellar population synthesis (SPS), makes use of these multi-wavelength (UV to IR) data to generate a library of model SEDs. The observed SEDs can then be compared with such a library using statistical fitting techniques like the Bayesian statistics to infer the physical properties of galaxies. The main focus of this thesis is on the reliability of stellar population synthesis modelling when only limited photometry of a small number of wavelength bands is available. It is divided in two parts: elaborate understanding of SPS modelling (\partref{part1}) and inferring galaxy physical properties (\partref{part2}).

    Some of the latest SPS modelling and SED fitting tools in the literature are the Code Investigating GALaxy Emission (CIGALE), Prospector, BAGPIPES and MAGPHYS. The general scheme is that these tools take the observed SED fluxes of galaxies as an input along with some model parameters depending on the specific tool, and output the inferred physical properties of those galaxies. Tools like CIGALE are very versatile and the flexibility comes at the cost of requiring a lot of user choices of models and parameter ranges. MAGPHYS has the advantage over many others that it is a highly user friendly and simple-to-use SPS modelling and SED fitting tool. The aim of this project is to reliably infer galaxy physical properties from a limited number of photometric bands. We chose to use MAGPHYS for this because of its user friendliness and proven reliability in other studies in the past (since 2008). We took the spectroscopic redshifts of galaxies from the ongoing large scale survey with the Dark Energy Spectroscopic Instrument (DESI) and the photometric data from the Legacy Surveys (DECaLS, BASS and MzLS). However, the catalog we used gives photometry of only the $g, r$ and $z$ optical bands and the $3.4\ \mu$m and the $4.6\ \mu$m (W1 and W2) bands of the Wide-field Infrared Survey Explorer (WISE), which is inadequate for reliable SPS modelling and SED fitting. Thus, we obtain the $12\ \mu$m and the $22\ \mu$m (W3 and W4) bands of WISE and the far-ultraviolet (FUV) and the near-ultraviolet (NUV) bands of the Galaxy Evolution Explorer (GALEX). To be able to compare our results with the past studies, we cross match our sample with the GALEX-SDSS-WISE Legacy Catalog (GSWLC)-X2. We first demonstrate SPS modelling and SED fitting on galaxies with panchromatic photometry data from the Galaxy and Mass Assembly (GAMA) \rnum{2} equatorial survey and then assess its reliability on galaxies with limited photometric coverage. Our investigations led us to conclude that the UV and the NIR fluxes are crucial for inferring the SFRs, the dust attenuation and emission by SPS modelling and SED fitting using MAGPHYS, and that the inferences are statistically reliable if there are at least 6 bands with a detection out of the 9 bands (FUV, NUV, $g, r, z$, W1, W2, W3 and W4).

    After a final verification of our approach, we plan to do the SPS modelling and SED fitting of the galaxies in the DESI Early Data Release and other future data releases.

%% file: 01_Chapters/ch01_intro.tex
\chapter{Introduction}\label{ch1}

    In astrophysical terms, a galaxy is defined to be a collection of a several millions to trillions of stars, their accompanying stellar systems, stellar remnants, the interstellar medium composed of gas, dust and dark matter, all bound together by gravity. Astrophysical processes in such a complex system usually generate electric and magnetic fields, and also radiate energy in the form of electromagnetic radiation. Apart from the advents of neutrino astronomy and gravitational wave astronomy, in most scenarios, an observer on the Earth, which itself is also a part of a galaxy we know by the name of the Milky Way galaxy, detects only the electromagnetic radiation that is emitted from a particular galaxy, or a collection of galaxies.
    
    The formation and evolution of such complex systems like galaxies and their collections, is an area of active research, with many questions lurking for definitive answers. Several developments have occurred both on the observational and theoretical/simulation fronts in recent times. Given the nature of the problem, one has to rely mostly on statistical studies of galaxy populations in the Universe and invoke ergodicity whenever required. In such studies, the physical properties of the galaxies, like galaxy morphology, local environment density, the stellar mass, the star formation rate (SFR), the star formation history (SFH), dust attenuation and emission parameters, metallicity, etc., are the key ingredients from which inferences on processes affecting galaxy evolution, are made. It well known that the various galaxy properties like morphology, stellar mass, gas properties and environment show correlations with each other, and it is not clear which of these are fundamental. (\cite{Bait17}). Thus, it is important and useful to infer galaxy physical properties.

    Galaxy morphology can be inferred from their images using machine learning or deep learning. Local environment density can be inferred from clustering studies in 3D (two sky coordinate and the redshift). Because the light of normal galaxies (those which do not have an active nucleus at the centre) originates from stars, the spectrum of galaxies can be interpreted as a superposition of stellar spectra. Stellar evolution is largely understood, and the spectral radiation of stars can be calculated from the theory of stellar atmospheres. We have also the observational spectral data to verify theoretical calculations. If the distribution of the number density of stars is known as a function of their mass, chemical composition, and evolutionary stage, we can compute the light emitted by them. We have to take into account also the fact that the distribution of stars changes over time, which means that the spectral distribution of the population also changes in time. The spectral energy distribution of a galaxy thus reflects its history of star formation and stellar evolution. The dust grains in the interstellar medium (ISM) absorbs UV photons and re-emits the absorbed energy in the form of IR photons. For this reason, comparing different simulated star formation histories and dust attenuation and emission models with observed galaxy spectra provides important clues for understanding the evolution of galaxies. Thus, many of the physical properties of galaxies can be inferred from their panchromatic spectral energy distribution (SED) using techniques like the stellar population synthesis (SPS) (\cite{Conroy13}), which forms the subject of the presented work. On the observational front, there has been a deluge of multi-wavelength data ranging from the ultra-violet to the radio, mainly due to wide field surveys e.g., the Sloan Digital Sky Survey (SDSS) and the Dark Energy Spectroscopic Instrument (DESI) survey in the optical. Such developments have also been coupled with recent developments on complete multi-wavelength spectral energy distribution (SED) modelling of galaxies. This has been possible due to a better understanding of stellar population synthesis and dust modelling. In this thesis, we make use of these large volumes of data and advanced SED modelling techniques like SPS to infer the physical properties of galaxies.
    
    \partref{part1} of this thesis presents an overview of the theory of stellar population synthesis. \chref{ch2} focuses on the so-called simple stellar populations (SSPs), and elaborating upon the ingredients required to generate them. \chref{ch3} focuses on the so-called composite stellar populations (CSPs), and elaborating upon the ingredients required to generate them. The \partref{part2} presents the statistical analysis for inferring galaxy physical parameters. \chref{ch4} discusses the implementation of SPS and the Bayesian analysis for interpreting galaxy SEDs in the code Multi-wavelength Analysis of Galaxy Physical Properties (MAGPHYS, \cite{daCunha08}). The thesis concludes with \textit{Results, Discussion,} and \textit{Future work} in \chref{ch5}.

%% file: 01_Chapters/ch02_SSP.tex
\chapter{Simple Stellar Populations}\label{ch2}
    \section{What are Simple Stellar Populations?}\label{sec2.1}
        Stellar population synthesis, as the name suggests, refers to the technique of synthesizing a library of different populations of stars. The usual purpose of SPS is to model the multi-wavelength SEDs of galaxies and to infer their physical properties using a statistical, model-fitting analysis. The starting point of any SPS model is a simple stellar population (SSP). It describes the evolution in time of the SED of a single, coeval stellar population at a single metallicity and abundance pattern. An SSP therefore requires three basic inputs: an initial mass function (IMF), stellar spectral libraries, and stellar evolution theory in the form of isochrones, each of which may in principle be a function of metallicity and/or elemental abundance pattern.
        
    \section{Initial Mass Function}\label{sec2.2}
\begin{figure}[h]
            \centering
            \includegraphics[scale=0.15]{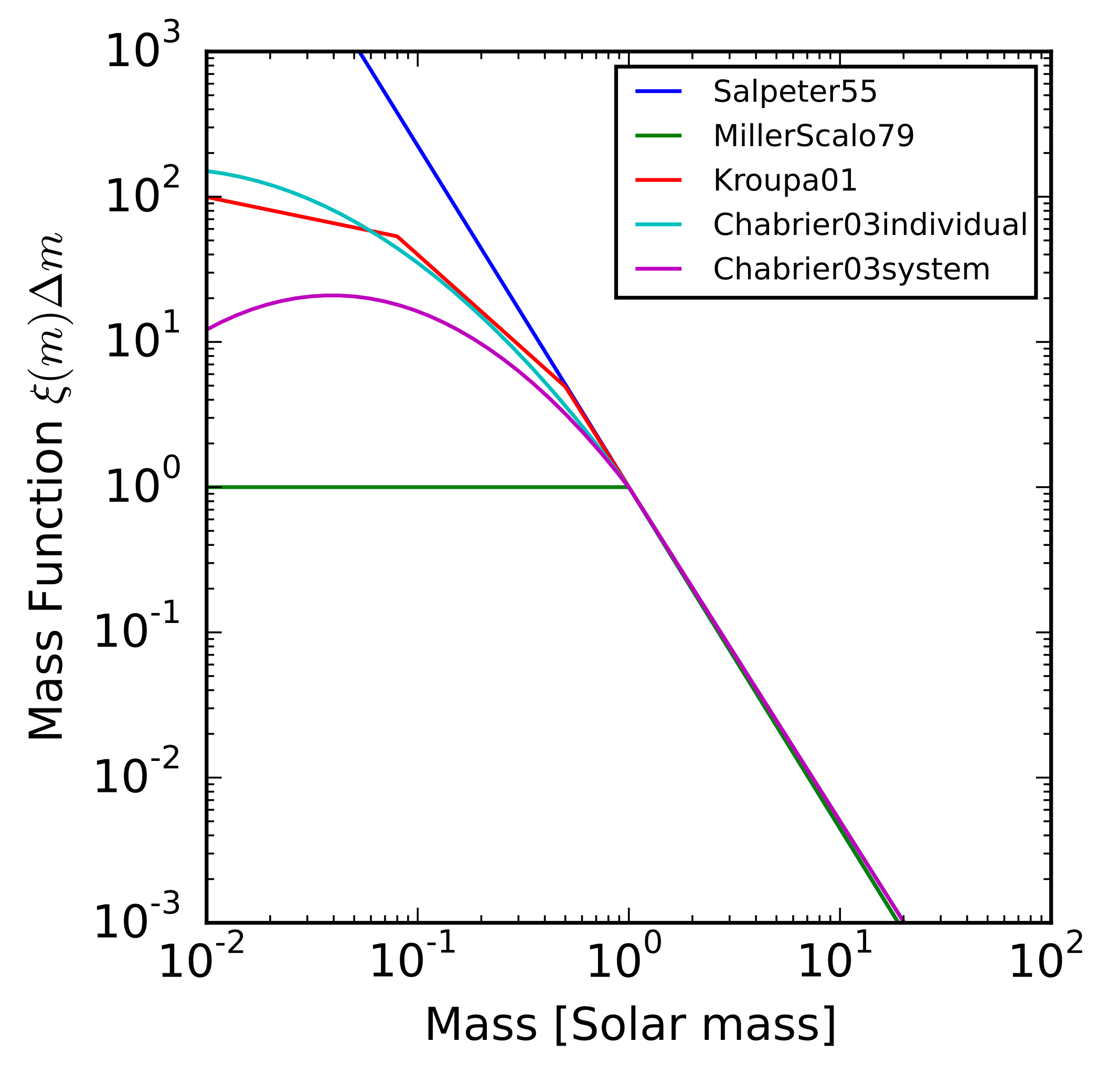}
            \caption{Various initial mass functions (IMFs) available in the literature: \cite{Salpeter55} (blue), \cite{MillerScalo79} (green), \cite{Kroupa01} (red) and \cite{Chabrier03} for individual stars (cyan) and binary or multiple star systems (magenta). Source: \href{https://en.wikipedia.org/wiki/Initial_mass_function}{Wikipedia: Initial mass function}}
            \label{fig:imf}
        \end{figure}

        The Universe is primarily made up of atomic hydrogen and helium gas, with lesser amounts of molecular hydrogen gas and trace amounts of other elements. In a typical gaseous nebula, which is also primarily composed of hydrogen and helium gas, the temperatures and densities are such that the self-gravitational force is balanced by the force due to kinetic pressure, which prevents it from collapsing. When such a gaseous cloud cools down by various cooling mechanisms, it collapses under its own gravity to form dense clumps of gas, which collapses further because of enhanced gravitational attraction in denser regions. As the gas collapses into denser fragments, the temperature inside the collapsing cores rises and eventually, it is high enough that nuclear fusion reactions get ignited. As hydrogen fuses into helium, it emits energy in the form of electromagnetic radiation, exerting radiation pressure on the collapsing gas, eventually halting the collapse in a dynamic equilibrium, and a star is born! A collapsing nebula may form several hundreds to millions of stars nearly simultaneously (time difference between the birth of stars being much smaller than the lifetime of the stars), depending on the size of the nebula and its internal and external dynamics. The mass distribution of the stars thus formed, at the time of their birth (zero-age main sequence; see \secref{sec2.4} for more details about stellar main sequence), can be described by an empirical function called the initial mass function (IMF). Several attempts have been made to express the IMF in an analytical form, and some of them are shown in \figref{fig:imf}). The currently best accepted IMF, and also the one used in the presented work, is the one given by Gilles Chabrier (\cite{Chabrier03}):
        \begin{equation}
            \xi(\rm{log}\ m) = 0.158 \times \rm{exp}\left[\frac{-(\rm{log}\ m - \rm{log}\ 0.08)^2}{2(0.69)^2}\right] \ pc^{-3} \cdot (\rm{log}\ M_\odot)^{-1} \ \rm{for} \ m<1
        \end{equation}
        
        In most scenarios, the IMF, to a good approximation, can be assumed to be an intensive property of a collapsing nebula - same for any cloud of gas in a galaxy and hence, for the entire galaxy itself. In special scenarios, especially in the presence of strong magnetic fields or turbulent interstellar winds, the IMF might differ significantly.
        
    \section{Stellar Spectra}\label{sec2.3}
\begin{figure}[h]
            \centering
            \includegraphics[scale=0.75]{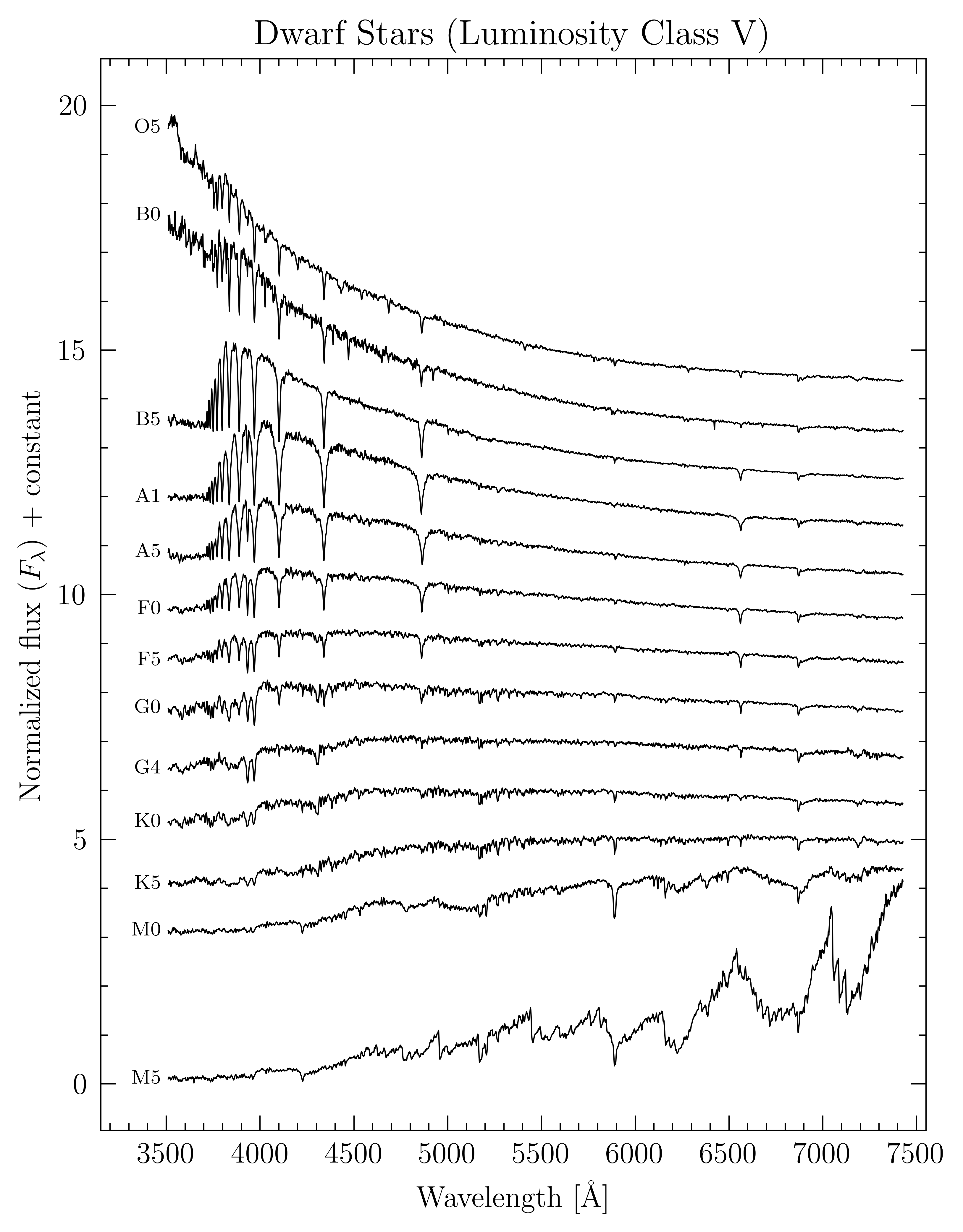}
            \caption{An illustration of the variety of stellar spectra of Type \rnum{5} dwarf stars. The original source spectra taken from \cite{Jacoby84} Source: \href{https://www.minerva.ufsc.br/~kanaan/intastro/lecture8/www.astronomy.ohio-state.edu/~pogge/Ast162/Unit1/SpTypes/index.html}{\copyright Richard W. Pogge}}
            \label{fig:stellar-spectra}
        \end{figure}

        As the nuclear reactions inside the core of a protostellar object, the surrounding gas, which eventually attains dynamical equilibrium under gravity to form the star, forms a medium of high optical depth and attain thermal equilibrium with the radiation emitted from the core. Thus, stellar spectra closely resembles that of a blackbody. However, a detailed stellar spectrum depends on the luminosity and surface temperature of the star, which, in-turn, are functions of its mass, composition and age. Stars can be classified into various spectroscopic categories depending on the relative strengths of various spectral lines (see \figref{fig:stellar-spectra}). The relative line ratios can be calculated from the equilibrium population of the energy levels, which requires detailed atomic modelling, and using the Einstein coefficients for transition lines, and the Saha ionization equation (\cite{Saha1920}) and Boltzmann's formula for the recombination lines. Since a wide variety of stars can form in a galaxy, detailed stellar spectral libraries are required for SPS modelling.
        
    \section{Isochrones}\label{sec2.4}
        As a star ages, synthesis of helium and other heavier elements and depletion of lithium due to various nuclear reactions in its core (see \cite{B2FH}, \cite{Baumann10}) changes its overall composition. Since a stellar spectrum depends on the composition of the star, the spectrum changes as the star ages due to the change in its composition. The finiteness of the lifespan of stars implies that the starlight contributed by the short-lived, hotter, bluer stars in a galaxy will diminish quicker than that from the relatively long-lived, cooler, redder ones, unless enough new star formation takes place. This also implies that older galaxies are redder than the relatively younger ones, unless there are gas inflows, fueling star formation or other starburst scenarios in the older ones or gas outflows or other quenching scenarios in the relatively younger ones. This is discussed in greater detail in the section on star formation histories, \secref{sec3.2}.
        
\begin{figure}[h]%
            \centering
            \subfloat[]{{\includegraphics[width=7cm]{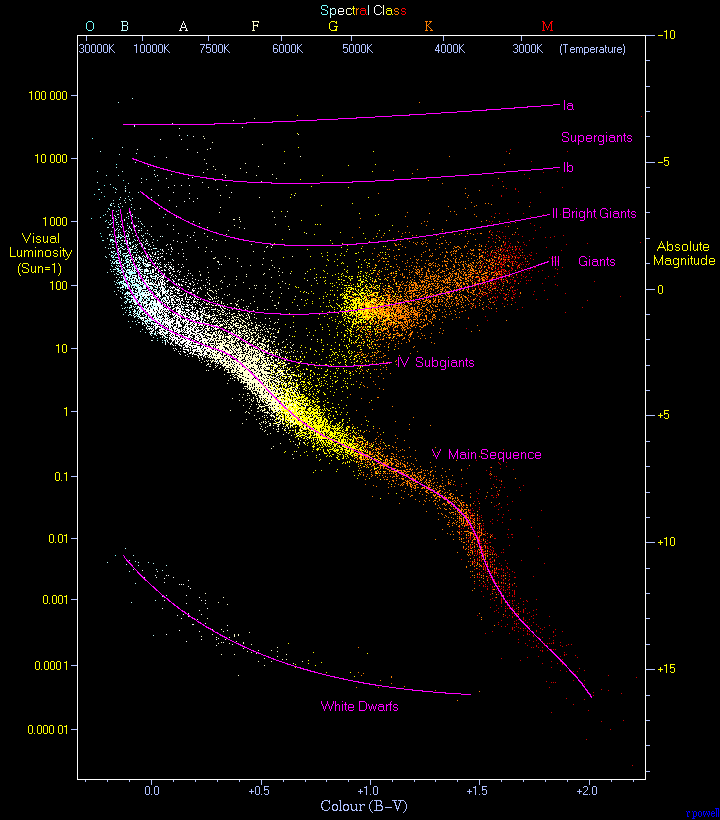}}}\\%
            \subfloat[]{{\includegraphics[width=5cm]{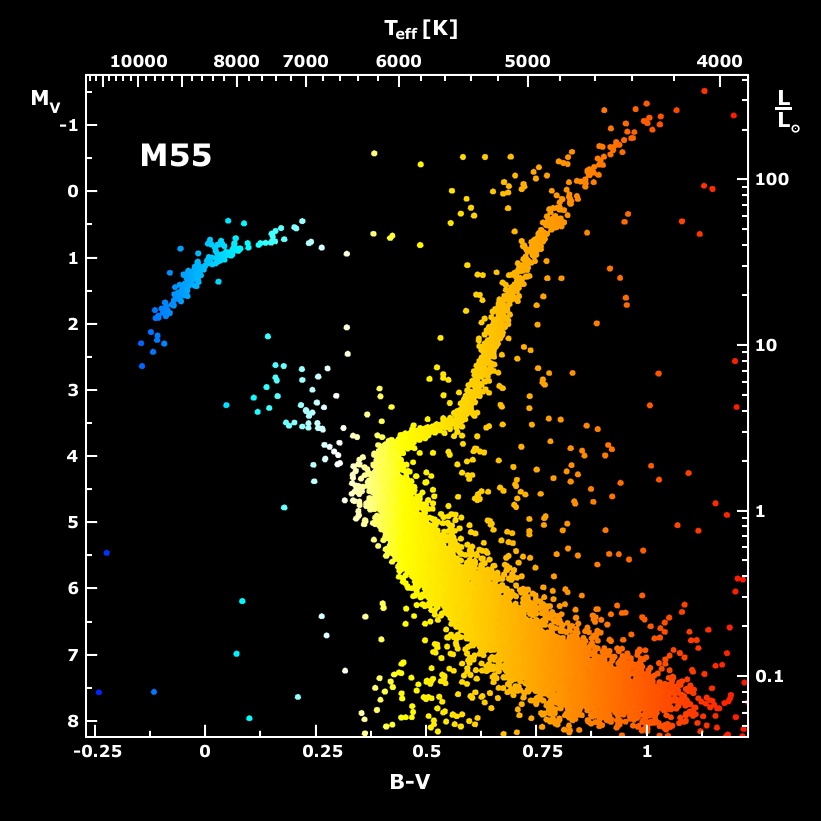}}}%
            \qquad
            \subfloat[]{{\includegraphics[width=5cm]{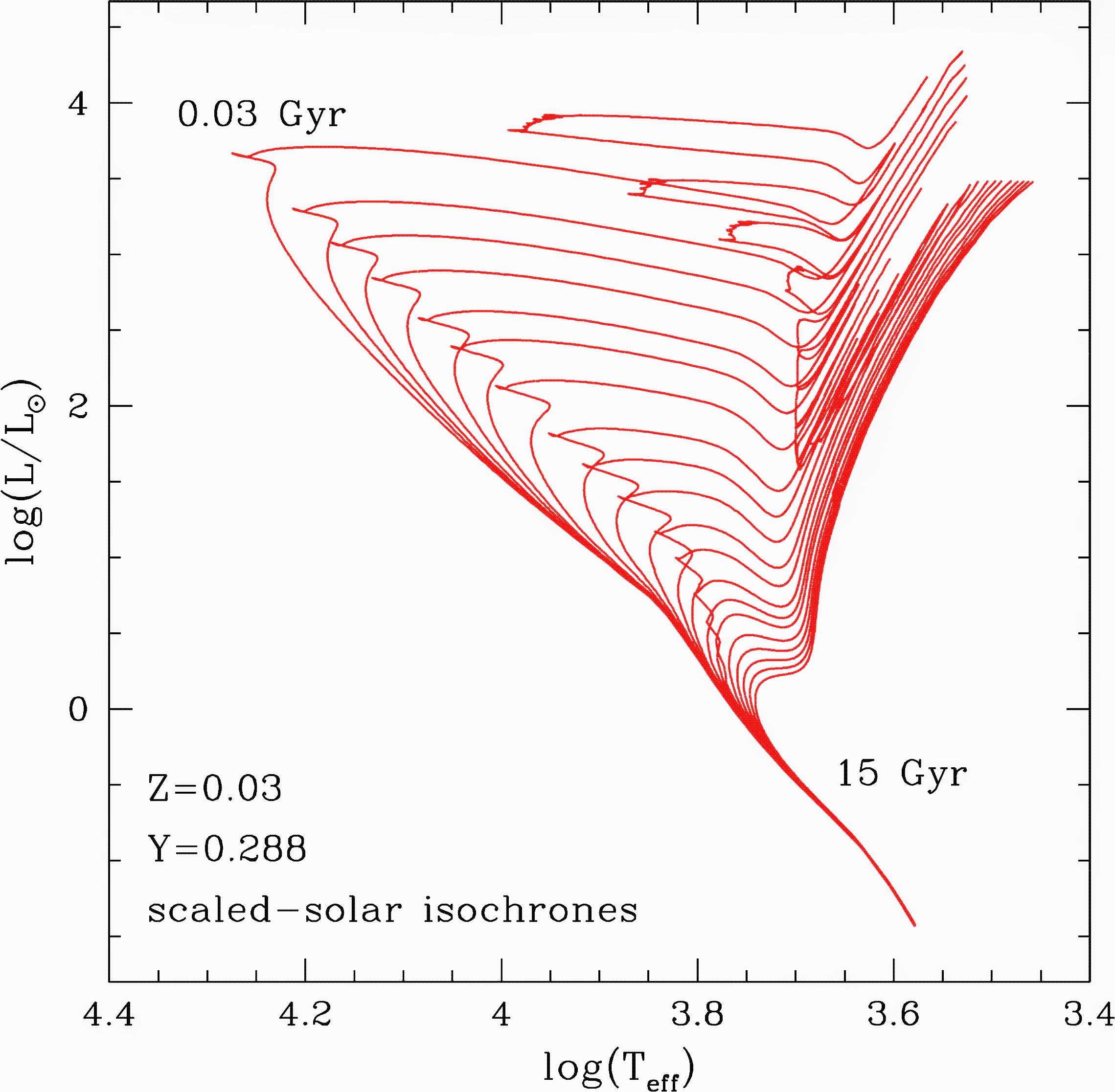} }}%
            \caption{Stellar evolution: HR diagram and stellar isochrone library. (a) An observational Hertzsprung–Russell diagram with 22,000 stars plotted from the Hipparcos Catalogue and 1,000 from the Gliese Catalogue of nearby stars. Source: \href{http://www.atlasoftheuniverse.com/hr.html}{\copyright Richard Powell} (b) HR Diagram of the Messier 55 (M55) globular star cluster. Source: \href{https://apod.nasa.gov/apod/ap010223.html}{NASA Astronomy Picture of the Day} (c) A sample selected from the updated Bag of Stellar Tracks and Isochrones (BaSTI) stellar isochrone library (\cite{BaSTI1}, \cite{BaSTI2}, \cite{BaSTI3}, \cite{BaSTI4}).}%
            \label{fig:hr-isochrones}%
        \end{figure}

        Stellar evolution can be understood largely from a plot of luminosity versus effective surface temperature for a number of stars, known as a Hertzsprung–Russell diagram (see \figref{fig:hr-isochrones} (a)), or an HR diagram in brief, after \cite{Hertzsprung1911} and \cite{Russell1913} who produced the first observational plots of this kind. Sometimes, other quantities like the absolute magnitude or the B-V color, which can be a proxy to either the luminosity or the effective surface temperature, are plotted in an HR diagram. In case of a star cluster, the interstellar distances are often negligible compared to their distances from the observer, and in such cases, plotting the apparent magnitudes gives essentially the same information about the stellar evolution as the absolute magnitudes or the luminosities. For historical reasons, the convention is to plot the effective surface temperature decreasing along the positive $x$-axis, the luminosity increasing along the positive $y$-axis, and the absolute magnitude decreasing along the positive $y$-axis. The nearly straight line of negative slope of about 6 in \figref{fig:hr-isochrones} (a) is called the stellar main sequence\footnote{It should not to be confused with the galaxy main sequence (see \cite{Gal_MS}).} On the main sequence, the stellar mass decreases along the positive $x$-axis. Since high mass stars have a shorter lifespan, in a population of stars born nearly simultaneously, like a globular cluster, the high mass stars near the left end of the HR diagram of that population of stars will evolve out of the main sequence earlier than other stars on the main sequence. As a result, there is a sharp turn-off from the main sequence (see \figref{fig:hr-isochrones} (b)). The age of the population is then the age of the stars at the turn-off point. As a star evolves off the main sequence, it might transform into a red giant star or a blue super giant star, and eventually a white dwarf, a neutron star or a black hole depending on its mass, which will have a different spectrum from the precursor main sequence star. Thus, post-main sequence evolution has a complex effect on the SSP of the concerned population of stars. A detailed understanding of stellar evolution results from extensive studies of composite HR diagrams of multiple stellar populations. This facilitates the prediction of the SSP of a stellar population at any given instant in time after its birth using isochrones. In essence, details of stellar evolution in the form of isochrones is required for SPS modelling.
        
    \section{Generating Simple Stellar Populations}\label{sec2.5}
\begin{figure}[h]
            \centering
            \includegraphics[scale=0.4]{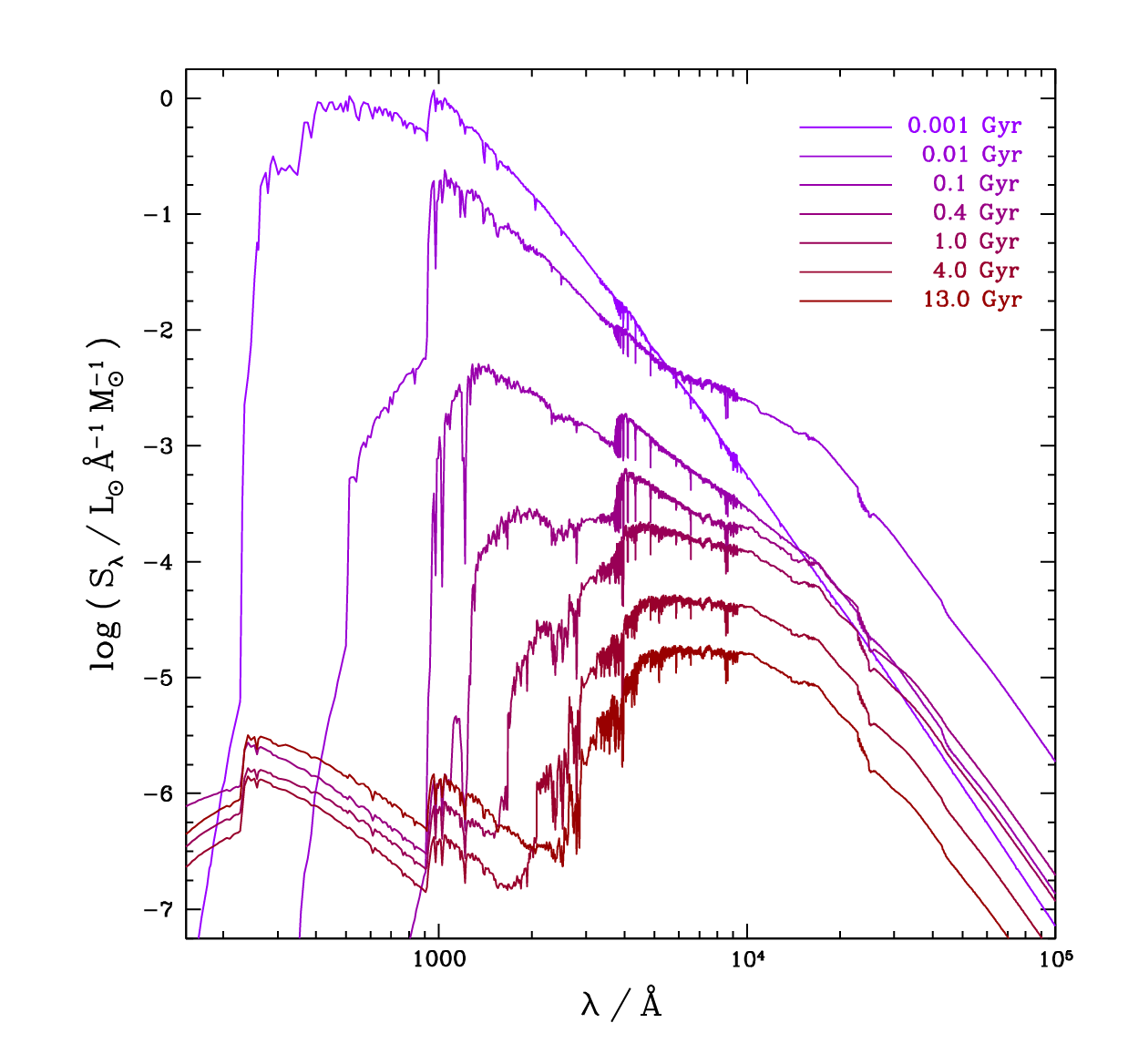}
            \caption{The simple stellar population library of \cite{BC03}. Source: \href{https://www.icrar.org/wp-content/uploads/2022/10/Talk-16-KF80_September2022.pdf}{\copyright Elisabete da Cunha}}
            \label{fig:ssp}
        \end{figure}

        To generate the SSPs (see \figref{fig:ssp}), the above ingredients are typically combined in the following way (\cite{Conroy13}):
        \begin{equation}
            f_{\rm{SSP}}(t,Z) = \int_{m_{\rm{lo}}}^{m_{\rm{up}}(t)} f_{\rm{star}}\left[T_{\rm{eff}}(M),\ \log g(M)|t,\ Z\right]\ \Phi(M)\ \rm{d}M,
        \end{equation}
        where $M$ is the zero-age main sequence stellar mass, $\Phi(M)$ is the IMF, $f_{\rm{star}}$ is a stellar spectrum, and $f_{\rm{SSP}}$ is the resulting time and metallicity-dependent SSP spectrum. The lower limit of integration, $m_{\rm{lo}}$ , is typically taken to be the hydrogen burning limit (either 0.08 or 0.1 $M_\odot$ , depending on the SPS code), and the upper limit is dictated by stellar evolution. The isochrones determine the relation between the effective surface temperature $(T_{\rm{eff}})$ , the logarithm of surface gravity $(\log\ g)$, and $M$ for a given time $(t)$ and metallicity $(Z)$.
        
%% file: 01_Chapters/ch03_CSP.tex
\chapter{Composite Stellar Populations}\label{ch3}
    \section{What are Composite Stellar Populations?}\label{sec3.1}
        Physical conditions in a typical galaxy are more complex than what is represented by simple stellar populations. SSPs lack many of the crucial astrophysical scenarios like star formation activities as a galaxy evolves, metallicity evolution of the galactic gas, and dust attenuation and emission. Thus, composite stellar populations (CSPs), which are more complex stellar systems than SSPs, differ from them in three respects: ($a$) they contain stars with a range of ages given by their SFH; ($b$) they contain stars with a range in metallicities as given by their time-dependent metallicity distribution function, $P(Z, t)$; and ($c$) they contain dust.
        
    \section{Star Formation History}\label{sec3.2}
        Stellar spectra closely resemble a blackbody spectrum of temperature equal to the effective surface temperature ($T_{\rm{eff}}$). Also, the $T_{\rm{eff}}$ of the main sequence stars lie in the range $10^5\ \rm{K}-10^3\ \rm{K}$, which implies that the a majority of the UV photons are contributed by the stars with a high $T_{\rm{eff}}$, i.e., the high-mass main sequence stars. Thus, the UV region of the spectrum of a stellar population is dominated by high-mass main sequence stars. Since high-mass main sequence stars have a shorter lifespan of around $10^8$ years, new star formation greatly affects the UV emission from a stellar population like a galaxy. The star formation rate (SFR) of a stellar population, averaged over the last $10^8$ years of its age, can thus be inferred from its UV flux. Other indicators of SFR include H$\alpha$, P$\alpha$, $24\rm{-}\mu \rm{m}$ luminosity, total IR luminosity, radio continuum luminosity, and even the X-ray ﬂux, and often a combination of several of these indicators is used (see, for example, \cite{Calzetti07}, \cite{Kennicutt07}).

\begin{figure}[h]
            \centering
            \includegraphics[scale=0.5]{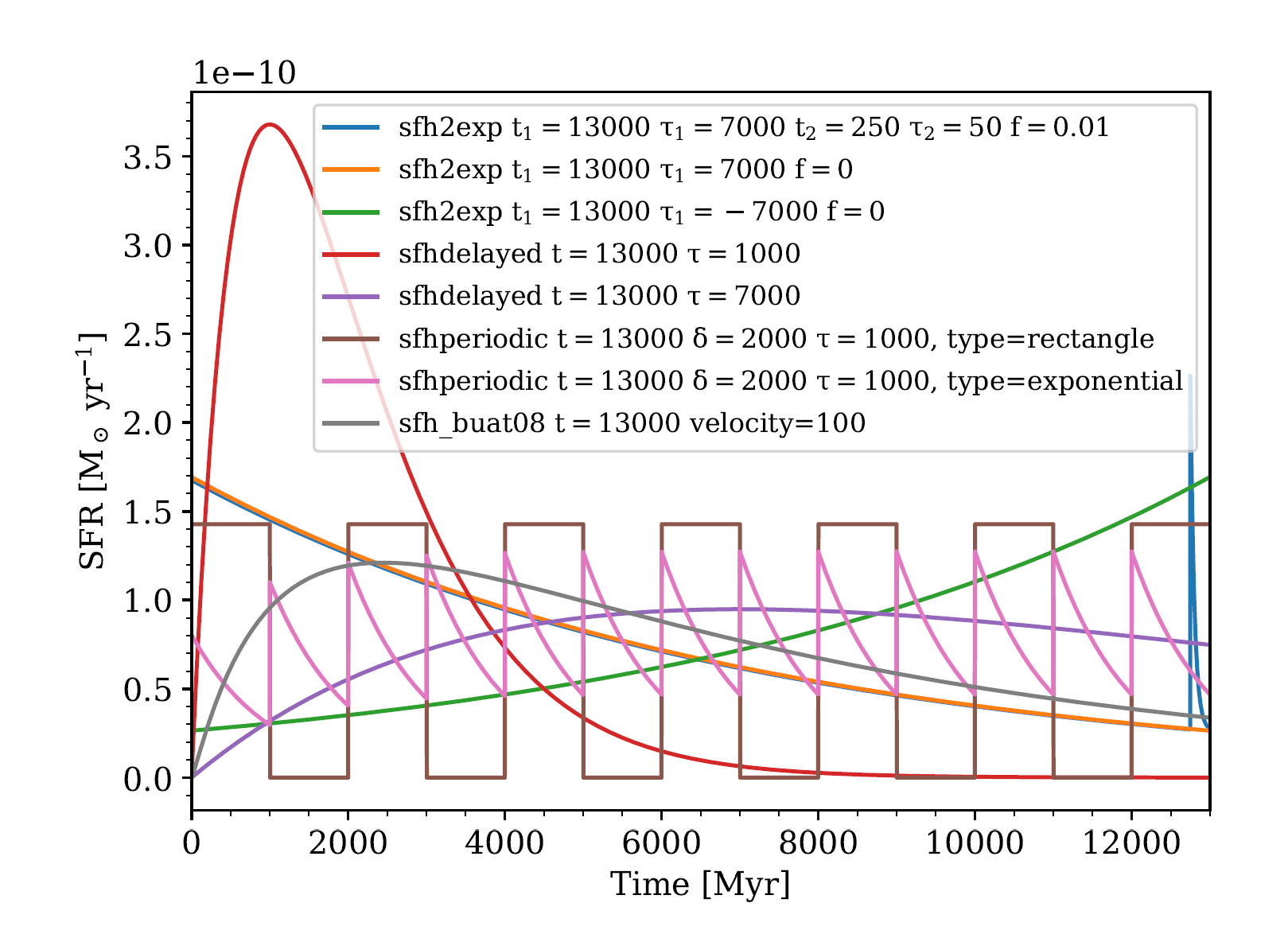}
            \caption{SFH models implemented in the Code Investigating GALaxy Emission (CIGALE, \cite{Boquien19}).}
            \label{fig:sfh}
        \end{figure}

        Star formation history (SFH) of an individual galaxy is impractically difficult to measure given the huge difference in evolution timescales of a galaxy and human lifetime. Thus, most SFH studies rely on the ergodic hypothesis, which asserts that that all accessible microstates are equiprobable over a long-enough period of time. Measuring the star formation rates in an ensemble of galaxies, one can make statistical inferences about the possible SFHs of galaxies. It is then possible to make informed guesses about analytical SFHs (see \figref{fig:sfh}) and by sampling the parameter space at near completeness, obtain a probability distribution function (PDF) for different SFHs for a galaxy.
        
    \section{Chemical Evolution}\label{sec3.3}
\begin{figure}[h]
            \centering
            \includegraphics[scale=0.7]{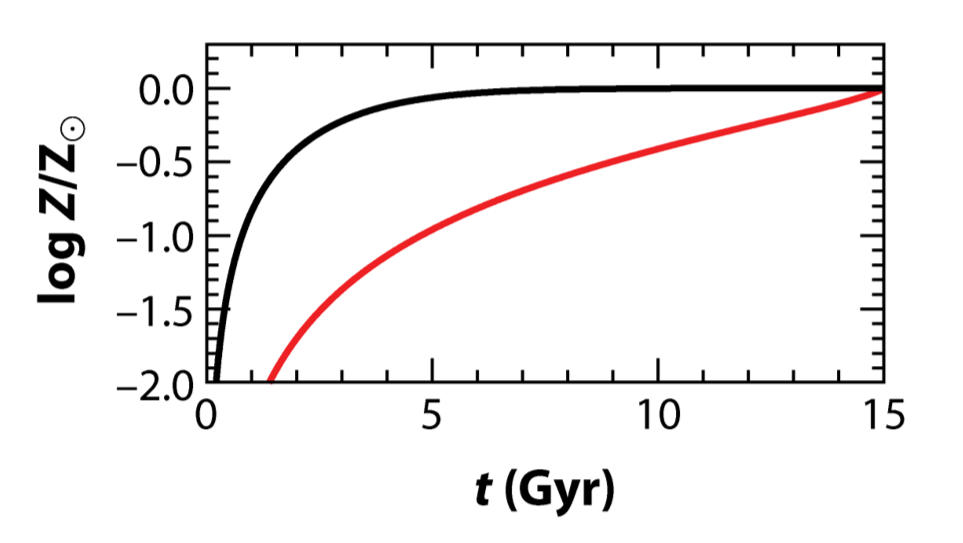}
            \caption{Illustrative examples of metallicity evolution models from \cite{Conroy13}.}
            \label{fig:metallicity_evol}
        \end{figure}

        According to the currently best accepted model of the Universe, it all started with a "Big Bang". Our current understanding of particle physics asserts that most of the nuclei other than hydrogen nuclei we see today were formed around 10 seconds to 20 minutes after the big bang, by a process referred to as the big bang nucleosynthesis. The Universe then was composed mainly of hydrogen and helium, along with small fractions of the hydrogen isotope deuterium ($^2$H or D), the helium isotope helium-3 ($^3$He), and a very small fraction of the lithium isotope lithium-7 ($^7$Li). Unstable isotopes like tritium ($^3$H or T) and the beryllium isotope beryllium-7 ($^7$Be) were also formed, but they later decayed into $^3$He and $^7$Li, respectively.  Elements heavier than lithium are thought to have been created later in the life of the Universe by stellar nucleosynthesis, through the formation, evolution and death of stars (\cite{B2FH}). Nuclear reactions in the core of a star leads to the formation of heavier nuclei of mass up to that of iron-56 ($^{56}$Fe). When the star dies, these heavier nuclei are ejected out, either by slowly blowing off the outer layers, forming a planetary nebula and a white dwarf, or in a cataclysmic explosion called a supernova, forming either a neutron star or a black hole, depending on the mass of the star and its core. In any case, a large fraction of the mass of the star is ejected out and mixed with the interstellar medium (ISM). Nuclei heavier than $^{56}$Fe are synthesised in supernovae explosions and mixed with the ISM. This is how metals (elements heavier than helium) enrich the ISM. To quantify the enrichment, metallicity ($Z$) can be used. It is defined as the abundance of iron:
        \begin{equation}
            Z\equiv\left[\frac{Fe}{H}\right]=\log\left(\frac{N_{Fe}}{N_H}\right)_*-\log\left(\frac{N_{Fe}}{N_H}\right)_\odot,
        \end{equation}
        where $N_{Fe}$ and $N_H$ are the number of iron and hydrogen atoms per unit of volume respectively, $\odot$ is the standard symbol for the Sun, and $\ast$ for a star. The metallicity of an isolated gas cloud increase with time, and so, later generations of stars are born with higher metallicities compared to the earlier ones. Since the SSPs has a metallicity and time dependence, evolution of metallicity implies evolution of the SSPs. Metallicity has been found to play a critical role for the dust mass growth by accreting materials in the interstellar medium (\cite{Asano13}). \cite{Yates12} have found correlation between metallicity, stellar mass and star formation in the Sloan Digital Sky Survey (SDSS) Data Release 7 (DR7)  galaxies. These studies might be useful in improving SPS modeling strategies. Thus, we need accurate metallicity evolution models (see \figref{fig:metallicity_evol}) as an input to SPS modelling codes.
        
    \section{Dust Attenuation}\label{sec3.4}
\begin{figure}[h]
            \centering
            \includegraphics[scale=0.8]{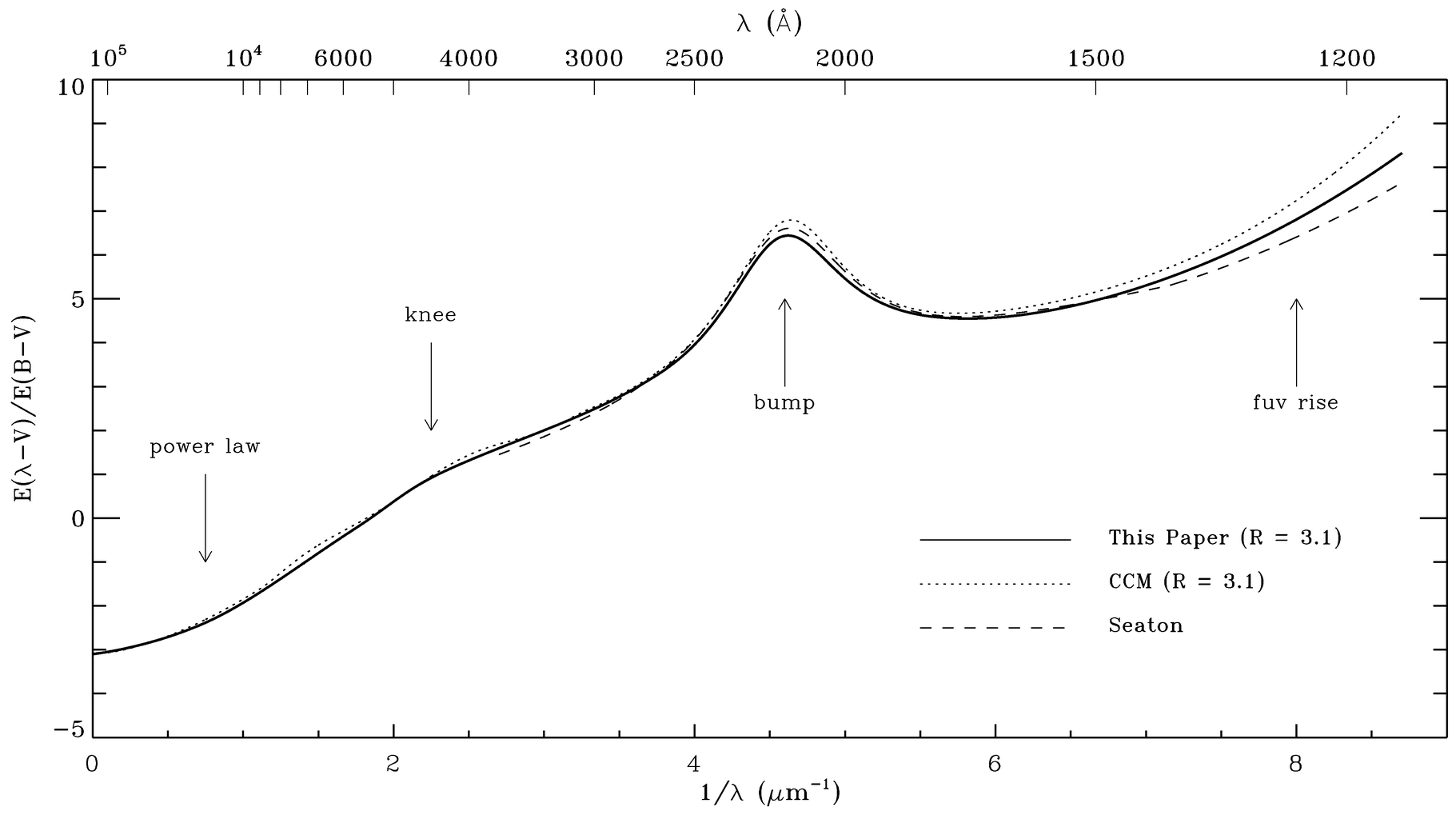}
            \caption{Dust attenuation law with various features marked. From \cite{Fitzpatrick99}, \cite{Cardelli89}, \cite{Seaton79}}
            \label{fig:dust_attenuation}
        \end{figure}

        Dust grains formed by various astrophysical processes in the ISM can attenuate the starlight from the far IR through the UV region. Typically, the extinction rises through the IR with a power-law–like dependence, rolls over slightly in the optical region (“knee”), shows a prominent feature at $2175\ \mbox{\AA}$ in the near-UV (“bump”), and has a sometimes steep rise in the far-UV (“fuv rise”) (\cite{Fitzpatrick99}) (see \figref{fig:dust_attenuation}). The extinction depends on the optical properties of the dust grains along a line of sight and potentially can reveal information about the composition and size distribution of the grains. There is no reason for the dust distribution to be uniform across a galaxy. Thus, changes in the extinction from place to place may reveal the degree and nature of dust grain processing occurring in the ISM. These can then be searched for association with stellar birth clouds, the ambient ISM (\cite{daCunha08}) or asymptotic giant branch stars (\cite{Conroy13}). Since the physical properties of dust grains in different environments will be different, its attenuation needs to be modelled differently. Such understanding gained from studies of the Milky Way galaxy can then be applied to model the dust attenuation in a more generic scenario of other galaxies. Since the UV photons are ones most affected by dust attenuation, the UV flux of a galaxy depends crucially on the dust content, which necessitates accurate dust modeling for SPS.
        
    \section{Dust Emission}\label{sec3.5}
\begin{figure}[h]
            \centering
            \includegraphics[scale=0.4]{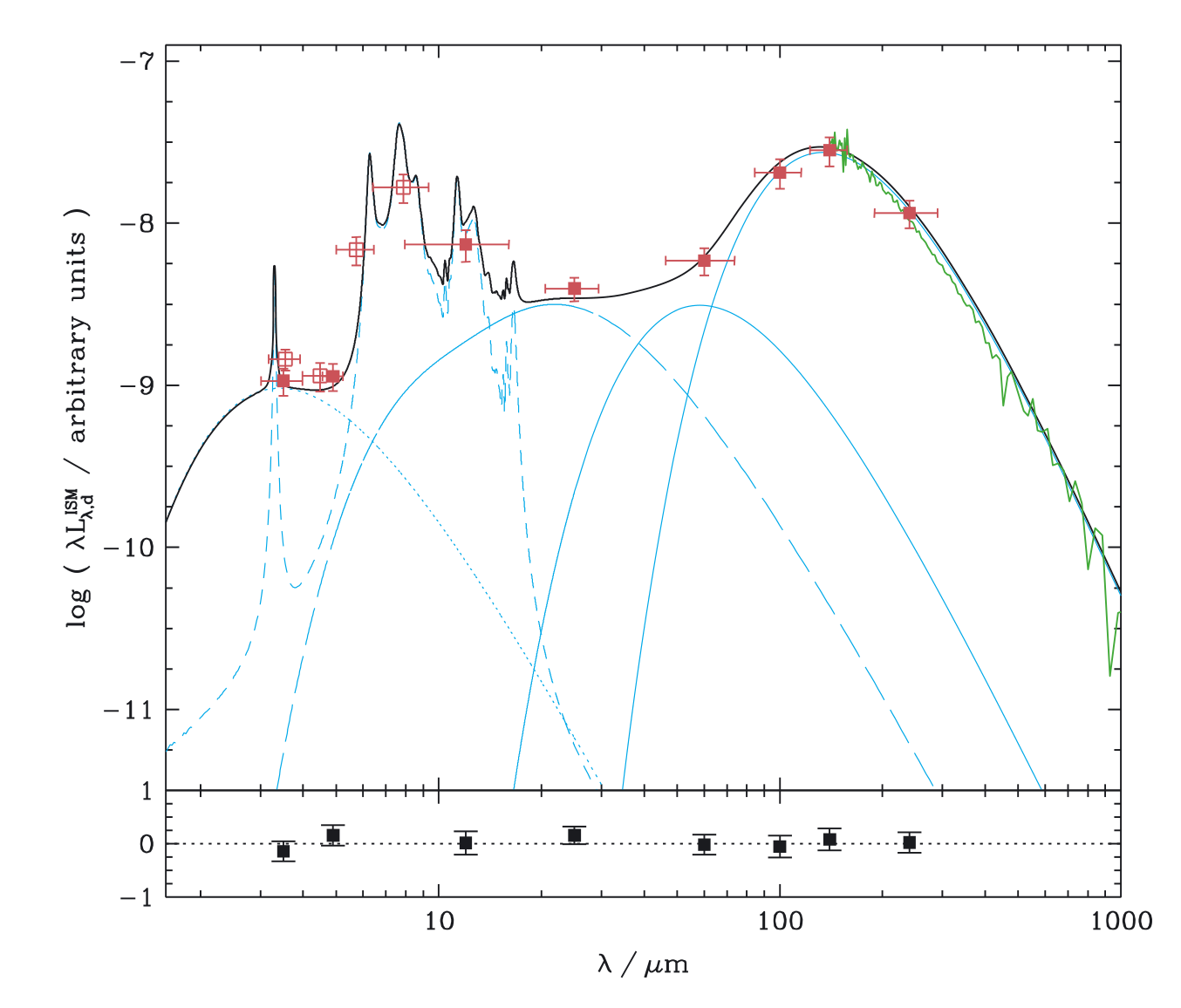}
            \caption{Dust emission model implemented in MAGPHYS \cite{daCunha08}. Best model fit (in black) to the observed mean spectral energy distribution of the Galactic cirrus emission. The red filled squares are the \textit{COBE}/DIRBE observations of \cite{Dwek97}. Also shown for reference are the \textit{Spitzer}/IRAC observations of \cite{Flagey06}, red open squares) and the \textit{COBE}/FIRAS observations of \cite{Dwek97}, green line). The blue lines show the decomposition of the model in its different components: near-infrared continuum (dotted); PAHs (short-dashed); hot mid-infrared continuum (long-dashed) and warm and cold grains in thermal equilibrium (solid). The fit residuals are shown at the bottom.}
            \label{fig:dust_emission}
        \end{figure}

        The energy absorbed by the dust grains in the form of UV photons is re-emitted at longer wavelengths from near-IR (NIR) to far-IR (FIR). Different components of dust contribute at different wavelengths. These contributions can broadly be divided into NIR continuum, emission from polycyclic aromatic hydrocarbons (PAH), mid-IR (MIR) continuum, and that from warm and cold dust in thermal equilibrium (\cite{daCunha08}) (see \figref{fig:dust_emission}). Details about the dust component abundances can thus be inferred from IR dust modelling. SFR can also be redundantly constrained from IR luminosity with accurate dust emission models because higher the SFR, higher will be the emitted and absorbed UV flux, which implies higher emission from dust grains. In essence, dust modelling is crucial to SPS.
        
    \section{Generating Composite Stellar Populations}\label{sec3.6}
\begin{figure}[h]
            \centering
            \includegraphics[scale=0.7]{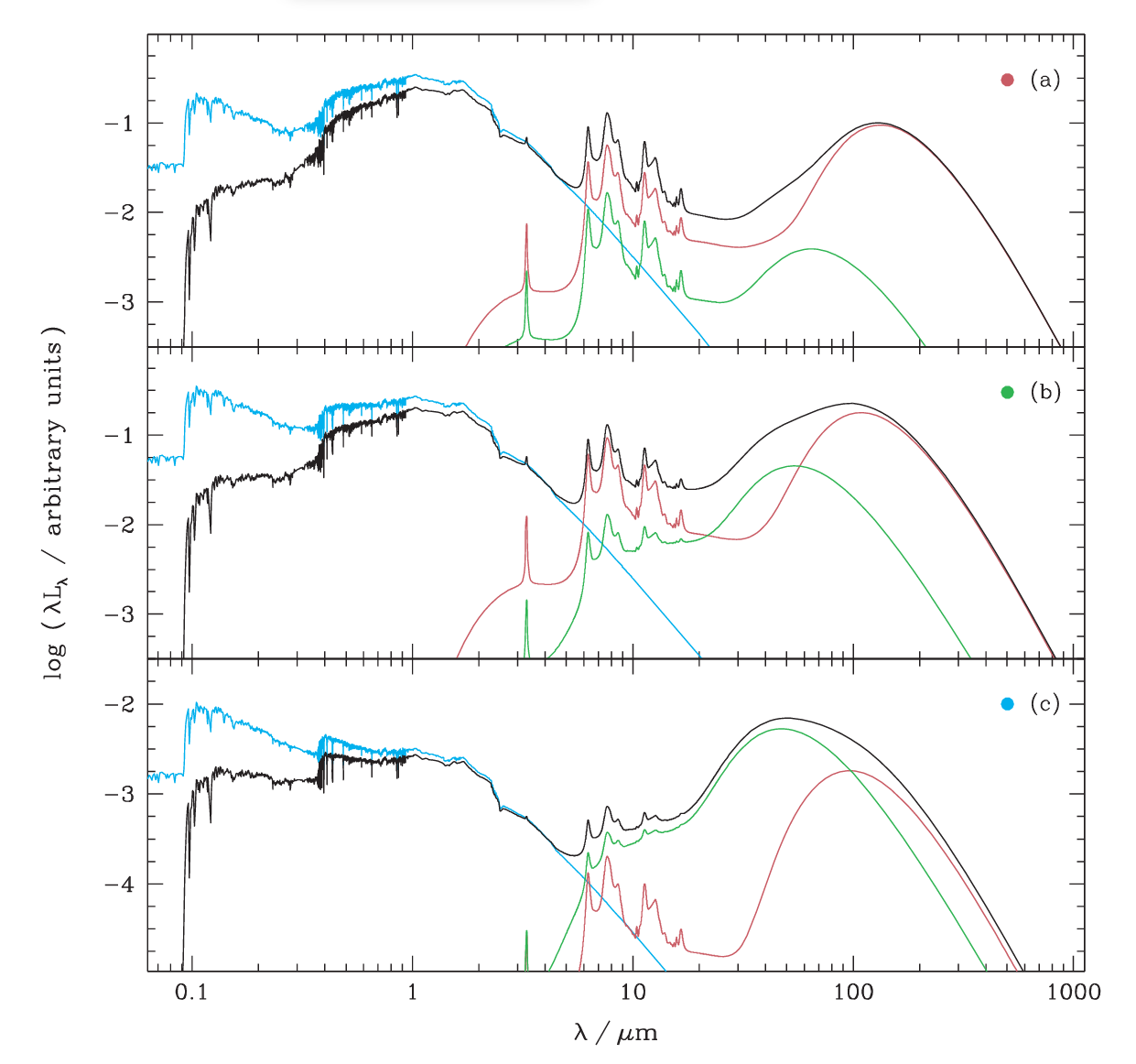}
            \caption{Example CSPs obtained using MAGPHYS \cite{daCunha08}, with various model components shown separately. The blue lines are the unattenuated CSPs and the black lines are the attenuated ones. (a) Quiescent star-forming galaxy; (b) normal star-forming galaxy; (c) starburst galaxy.}
            \label{fig:csp}
        \end{figure}

        To generate the CSPs, the above ingredients are typically combined in the following way (\cite{Conroy13}):
        \begin{equation}
            f_{\rm{CSP}}(t) = \int_{t'=0}^{t'=t}\int_{Z=0}^{Z_{\rm{max}}} (\rm{SFR}(t-t')P(Z,t-t')f_{\rm{SSP}}(t', Z)e^{-\tau_d(t')}+A\ f_{\rm{dust}}(t',Z))\ \rm{d}t'\rm{d}Z,
        \end{equation}
        where the integration variables are the stellar population age, $t'$, and metallicity, $Z$. Time-dependent dust attenuation is modeled via the dust optical depth, $\tau_d(t')$, and dust emission is incorporated in the function $f_{\rm{dust}}$. The normalization constant $A$ is set by balancing the luminosity absorbed by dust with the total luminosity re-radiated by dust. Full SED models from far-UV to far-IR can thus be obtained (see \figref{fig:csp}).
        
    \section{Summary of Stellar Population Synthesis}\label{sec3.7}
\begin{figure}
            \centering
            \includegraphics{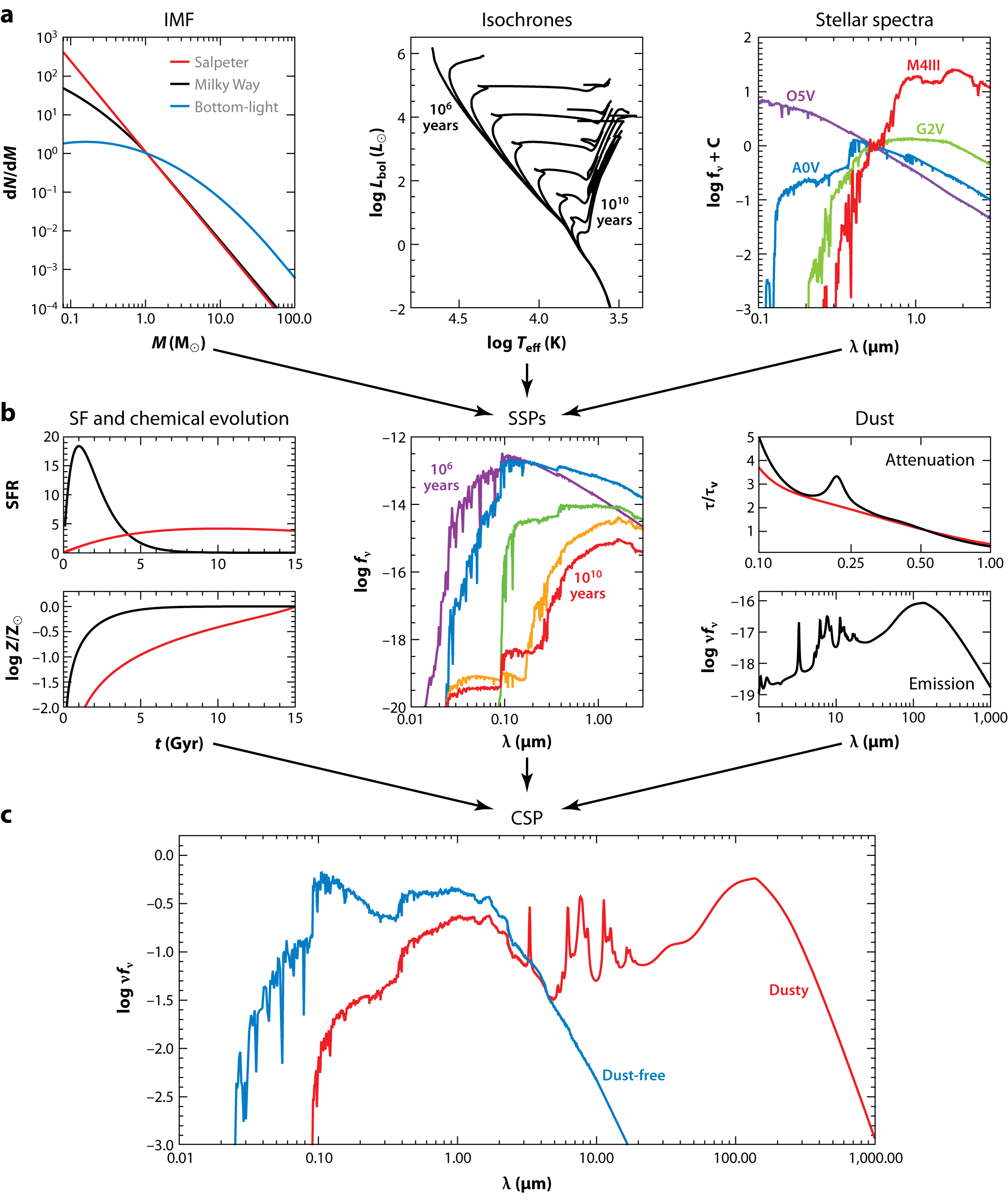}
            \caption{An overview of the SPS modelling procedure highlighting different stages along with the ingredients used. Source: \cite{Conroy13}}
            \label{fig:sps_overview}
        \end{figure}

        \figref{fig:sps_overview} presents an overview of the stellar population synthesis (SPS) modelling procedure. The process of SPS can be broken into two steps: generating simple stellar populations (SSPs), and generating composite stellar populations (CSPs). The ingredients required for generating SSPs are the IMF, a library of stellar spectra, and the isochrones. These combined gives the SSPs which further require star formation histories, metallicity evolution models, and dust attenuation and emission models to generate the CSPs. The modelled spectrum thus obtained is a function of the age of the synthesised stellar population.
        
%% file: 01_Chapters/ch04_MAGPHYS.tex
\chapter{Multi-wavelength Analysis of Galaxy PHYSical Properties}\label{ch4}
    \section{Stellar Population Synthesis Modelling}\label{sec4.1}
        Over the past several decades, considerable effort has been devoted to model the SEDs of galaxies from FUV to FIR. These have resulted in numerous state-of-the-art SPS modelling and SED fitting tools. Among the most recently developed ones, to cite a few of them, are the Code for Investigating GALaxy Emission (\texttt{CIGALE}, \cite{Boquien19}), the Multi-wavelength Analysis of Galaxy PHYSical Properties (\texttt{MAGPHYS}, \cite{daCunha08}), Prospector (\cite{Prospector1}, \cite{Prospector2}), \texttt{BAGPIPES} (\cite{BAGPIPES}) and the GRAphite-SILicate approach (\texttt{GRASIL}, \cite{GRASIL}). \cite{Pacifici23} have done a comprehensive comparative analysis of the output of 14 SED-fitting codes applied to the same observational photometric data set. We choose to use the MAGPHYS (original package version) code because of its simplicity, flexible dust modelling and user-friendliness.

        MAGPHYS provides a simple model to interpret the ultraviolet, optical and infrared emission from galaxies. It is largely an empirical but physically motivated model for SPS. One of its main assumptions is the conservation of the energy absorbed and reradiated by dust. However, one of its shortcomings is that it assumes that the dust in the ISM of galaxies is heated only by starlight, ignoring the possible influence of an active galactic nucleus. Numerous models for the ingredients of SPS exists in the literature. Among the most popular IMFs are the ones by \cite{Salpeter55}, \cite{MillerScalo79}, \cite{Kroupa01} and \cite{Chabrier03}. The Modules for Experiments in Stellar Astrophysics (MESA) Isochrones and Stellar Tracks (MIST) project (\cite{MESA}, \cite{MIST}) and the updated Bag of Stellar Tracks and Isochrones (BaSTI) stellar model and isochrone library (\cite{BaSTI1}, \cite{BaSTI2}, \cite{BaSTI3}, \cite{BaSTI4}) are among the most extensive and elaborate isochrones libraries. Large and comprehensive empirical stellar spectral libraries like the MaNGA Stellar Library (SDSS-IV MaStar, \cite{Yan19}) and the Medium-resolution Isaac Newton Telescope Library of Empirical Spectra (\texttt{MILES}, \cite{MILES}, \cite{MILES_up}), are now available to enable more reliable SSP modelling. Among the most popular libraries of SSPs are that of \cite{BC03} and \cite{Maraston05}. Numerous analytical models for the SFH exists in the literature (see, for example, \cite{Boquien19}, \cite{Buat08}, \cite{BC03}). MAGPHYS uses the most recent version of the \cite{BC03} SPS code which predicts the spectral evolution of stellar populations at wavelengths from $91\ \mbox{\AA}$ to $160\ \mu$m and at ages between $1 \times 10^5$ and $2 \times 10^{10}$ yr, for different metallicities (0.0001, 0.0004, 0.004, 0.008, 0.02, and 0.05), IMFs (\cite{Salpeter55} and \cite{Chabrier03}) and SFHs. It also incorporates a new prescription by \cite{Marigo07} for the thermally pulsating asymptotic giant branch (TP-AGB) evolution of low- and intermediate-mass stars. Numerous studies have focused on determining attenuation laws in galaxies, finding a remarkable diversity (for example, \cite{Calzetti00}, \cite{Wild11}, \cite{Reddy15}, \cite{Reddy16}, \cite{LoFaro17}, \cite{Buat18}, \cite{Salim18}). MAGPHYS compute the attenuation of starlight by dust using the simple, angle-averaged model of \cite{CF00}. This accounts for the fact that stars are born in dense molecular clouds, which dissipate typically on a time-scale $t_0 \sim 10^7$ yr. Thus, the emission from stars younger than $t_0$ is more attenuated than that from older stars. The modelling of dust emission is a very active domain of research, building on several generations of increasingly more powerful IR instruments, from IRAS to Herschel. Some of the most widely accepted dust emission models are \cite{Draine07}, \cite{Draine14}, \cite{Dale14}, \cite{Casey12} and the \texttt{THEMIS} dust model (\cite{THEMIS}) developed by the DustPedia team. The authors of MAGPHYS developed a simple but physically motivated prescription to compute the spectral distribution of the energy re-radiated by dust in the infrared (\cite{daCunha08}). These models are then combined as discussed in \partref{part1} to generate a library of CSPs at different redshifts of interest.
        
    \section{Spectral Energy Distribution  Fitting}\label{sec4.2}
\begin{figure}[h]
            \centering
            \includegraphics[scale=0.73]{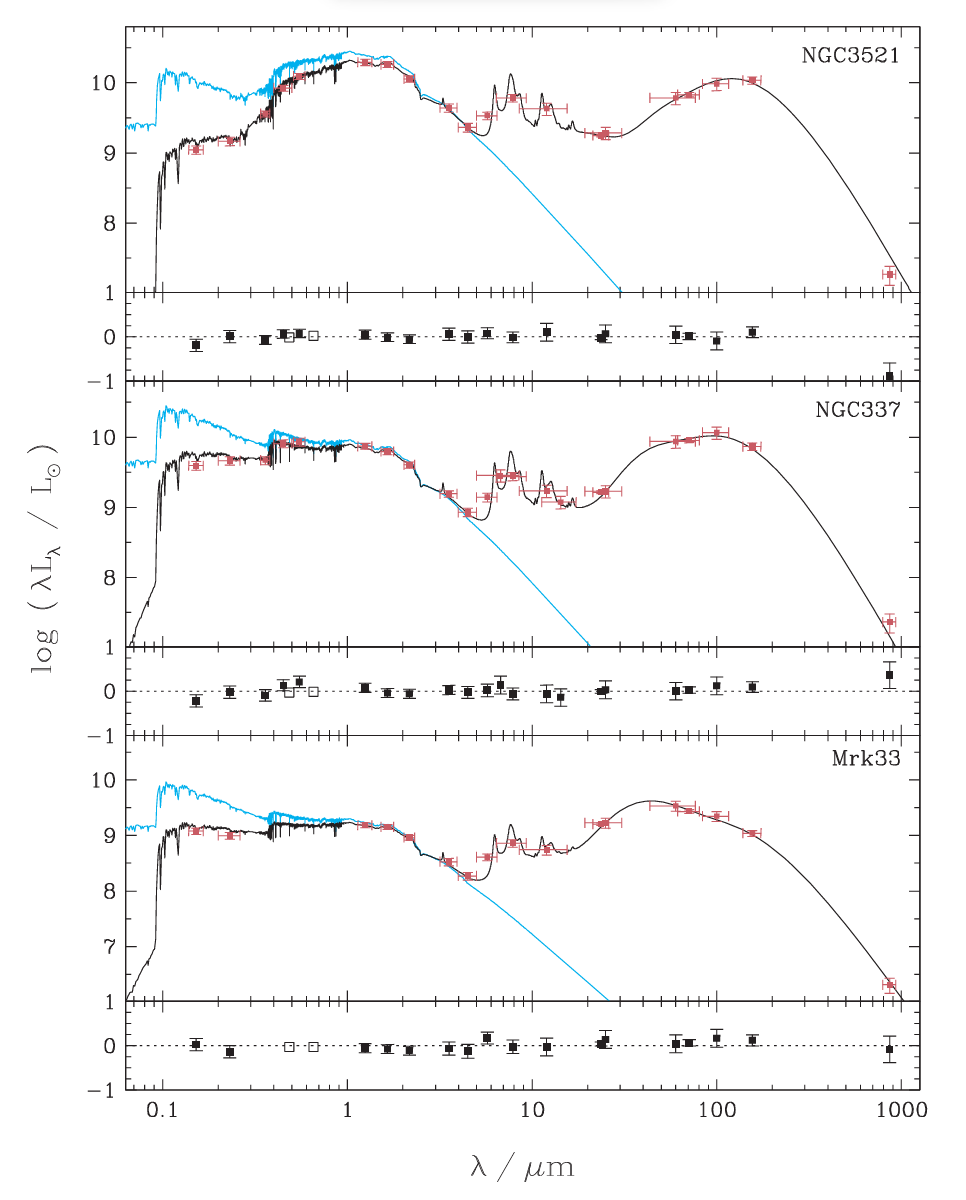}
            \caption{Best fit models (in black) to the observed spectral energy distributions (in red) of the three galaxies NGC 3521 (top panel), NGC 337 (middle panel) and Mrk 33 (bottom panel). In each panel, the blue line shows the unattenuated stellar population spectrum. For each observational point, the vertical error bar indicates the measurement error, while the horizontal error bar shows the effective width of the corresponding photometric bands. The residuals are shown at the bottom of each panel. From \cite{daCunha08}}
            \label{fig:sed_fitting}
        \end{figure}

        MAGPHYS takes an input file with the first column for a unique identification number for the galaxies in the file, the second column for their respective redshifts, and the rest of the columns in pairs of two for their respective photometric fluxes and flux errors in Jansky (Jy), and each row in the file corresponds to a galaxy. The fluxes form the SED of an observed galaxy in the sample. It performs a Bayesian analysis to generate the probability distribution functions (PDFs) of various physical properties of the galaxies. It also gives the parameters corresponding to the best fit models and the best fit unattenuated and attenuated SEDs - the quantity $\lambda L_\lambda / L_\odot$, where $L_\lambda$ is the specific luminosity (per unit wavelength) at wavelength $\lambda$ in units of the luminosity of the Sun ($L_\odot$) (see \figref{fig:sed_fitting}).
        
%% file: 01_Chapters/ch05_conclusion.tex
\chapter{Results and Conclusion}\label{ch5}
    \section{Sample Selection}\label{sec5.1}
        We intend to do SPS modelling of the galaxies in the Dark Energy Spectroscopic Instrument (DESI) (\cite{DESI}) spectroscopic survey. The DESI redshift value added catalog\footnote{DESI redshift VAC is hosted at \url{https://data.desi.lbl.gov/public/edr/vac/edr/zcat/fuji/v1.0/zall-pix-edr-vac.fits}} from the Early Data Release (EDR) (\cite{DESI_EDR}) provides photometry measurements from the Data Release 9 of the Legacy Surveys (\cite{LS_DR9_2019}) and the All-Sky Data Release of the Wide-field Infrared Survey Explorer (WISE) (\cite{WISE}) along with spectroscopic redshifts of the observed targets. It provides photometry measurements in the $g,\ r$ and $z$ bands of the Dark Energy Camera Legacy Survey (DECaLS) (\cite{LS_DR9_2019}), the $g$ and $r$ bands of the Beijing Arizona Sky Survey (BASS) (\cite{BASS}), the $z$ band of the Mayall z-band Legacy Survey (MzLS) (\cite{LS_DR9_2019}) and the $3.4\ \mu$m and the $4.6\ \mu$m (W1 and W2) bands of WISE. It has about 1.8 million objects. On this catalog, we impose the selection criteria of ($\texttt{SURVEY}==``\rm{sv3}"\ \&\&\ \texttt{ZWARN}==0\ \&\&\ \texttt{SPECTYPE}==``\rm{GALAXY}"\ \&\&\ \texttt{ZCAT\_PRIMARY}\ \&\&\ \texttt{PHOTSYS}\ !=\ ``\rm{G}"$), which leaves $6,84,839$ galaxies in the sample. The $5$ bands mentioned above does not provide adequate wavelength coverage for reliable SED modelling using SPS. So, we obtained the WISE $12\ \mu$m and $22\ \mu$m (W3 and W4) fluxes from the AGN Host Galaxies Physical Properties value added catalog from the DESI-EDR and cross-matched the above sample with the latest data release of Galaxy Evolution Explorer (GALEX) (\cite{GALEX}) to obtain the far-UV (FUV) and the near-UV (NUV) fluxes. This forms a sample of $2,97,402$ galaxies (referred to as the DESI-EDR sample henceforth). We then select a sub-sample cross matched with the GALEX-SDSS-WISE Legacy Catalog (GSWLC)-X2\footnote{GSWLC-X2 is hosted at \url{https://salims.pages.iu.edu/gswlc/GSWLC-X2.dat.gz}} (\cite{Salim16}, \cite{Salim18}), in which the authors did SPS modelling of the galaxies in the SDSS (\cite{SDSS}) using the CIGALE (\cite{Boquien19}), in order to be able to compare our results with those from past studies. Before doing the SPS modelling of this sample of $11,574$ galaxies with limited photometry (referred to as the DESI-GSWLC sample henceforth), we demonstrate it on the galaxies with panchromatic photometry data from the $\texttt{21BandPhotom}$ table\footnote{The $\texttt{21BandPhotom}$ table is hosted at \url{https://www.gama-survey.org/dr4/data/cat/PanchromaticPhotom/v03/21BandPhotomv03.fits}} from the PanchromaticPhotom DMU of the Galaxy and Mass Assembly (GAMA) \rnum{2} equatorial survey (\cite{GAMA}). The $\texttt{MagPhysv06}$ table\footnote{The $\texttt{MagPhysv06}$ table is hosted at \url{https://www.gama-survey.org/dr4/data/cat/MagPhys/v06/MagPhysv06.fits}} in the MagPhys DMU of the GAMA \rnum{2} equatorial survey provides physical stellar population and ISM parameters for galaxies in the GAMA \rnum{2} equatorial survey regions, derived by running the SED-fitting code MAGPHYS (\cite{daCunha08}) on $\texttt{LambdarCatv01}$\footnote{The $\texttt{LambdarCatv01}$ table is hosted at \url{http://www.gama-survey.org/dr4/data/cat/LambdarPhotometry/v01/LambdarCatv01.fits}}, a table in the LambdarPhotometry DMU in GAMA \rnum{2} (\cite{LambdarPhotom}). The sample obtained after cross-matching with $\texttt{GSWLC-X2}$ was then cross-matched with the $\texttt{21BandPhotom}$ and the $\texttt{MagPhysv06}$ tables to obtain a sample of $4,391$ galaxies (referred to as the DESI-21Band sample henceforth).
    
    \section{Results and Discussion}\label{sec5.2}
\begin{figure}%
            \centering
            \subfloat{{\includegraphics[width=17cm]{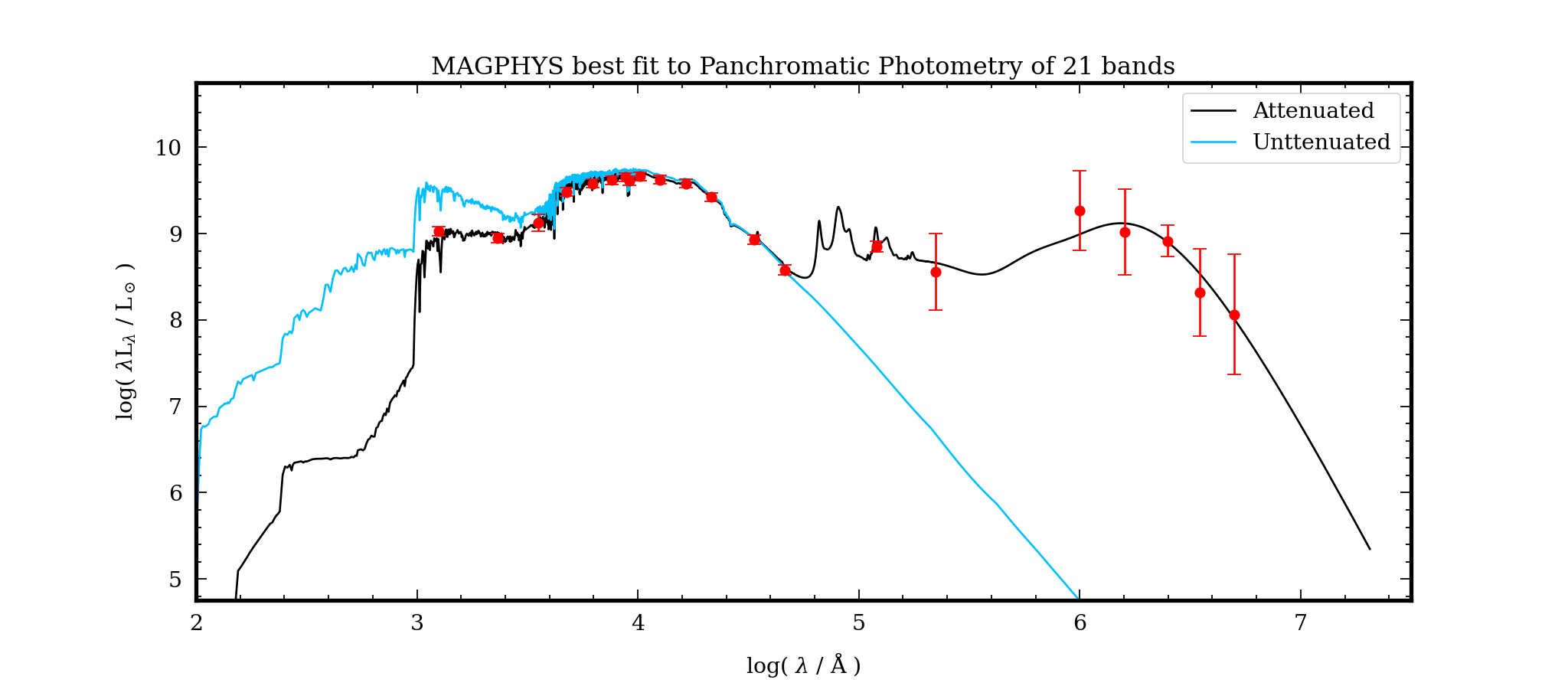}}}\\%
            \subfloat{{\includegraphics[width=17cm]{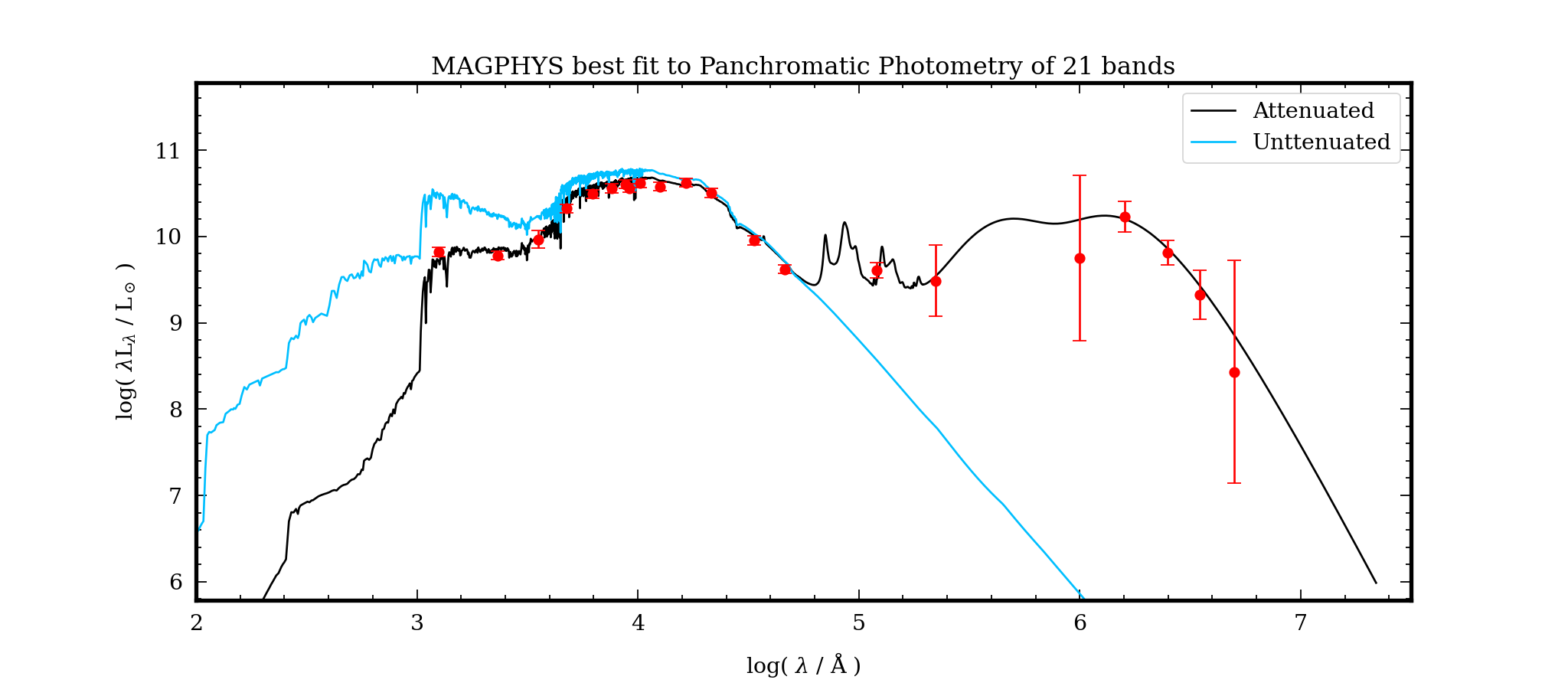}}}%
            \caption{Best fit models (in black) to panchromatic photometry of 21 bands (in red) of two randomly selected galaxies from the DESI-21Band sample. In each panel, the blue line shows the unattenuated stellar population spectrum. For each observational point, the error bars indicates the measurement error.}%
            \label{fig:21band-csp}%
        \end{figure}

        We first demonstrate SPS modelling and SED fitting of the galaxies in the DESI-21Band sample using MAGPHYS \cite{daCunha08}. \figref{fig:21band-csp} shows the best fit CSPs of two randomly selected galaxies from the DESI-21Band sample. The best fit CSPs (attenuated, black) are consistent with the observed SED (red circles) given the uncertainty in the observed fluxes (converted to luminosities) as represented by the errorbars. Here, the large errorbars at longer wavelengths indicate lower signal-to-noise ratio (SNR) for those photometric measurements. The agreement is similar to that obtained by the authors of MAGPHYS (see \figref{fig:sed_fitting}).
        
\begin{figure}%
            \centering
            \subfloat[]{{\includegraphics[width=11cm]{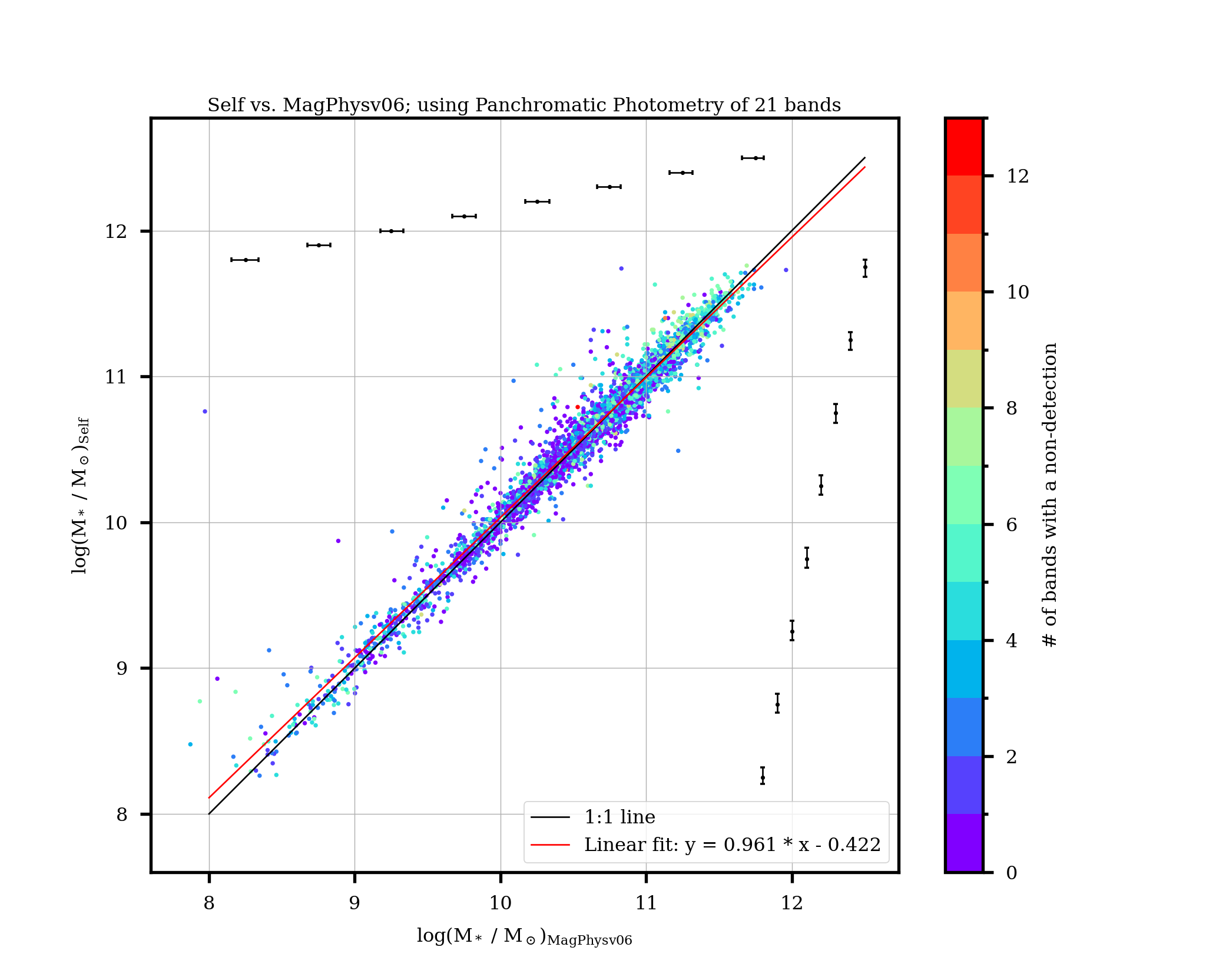}}}\\%
            \subfloat[]{{\includegraphics[width=11cm]{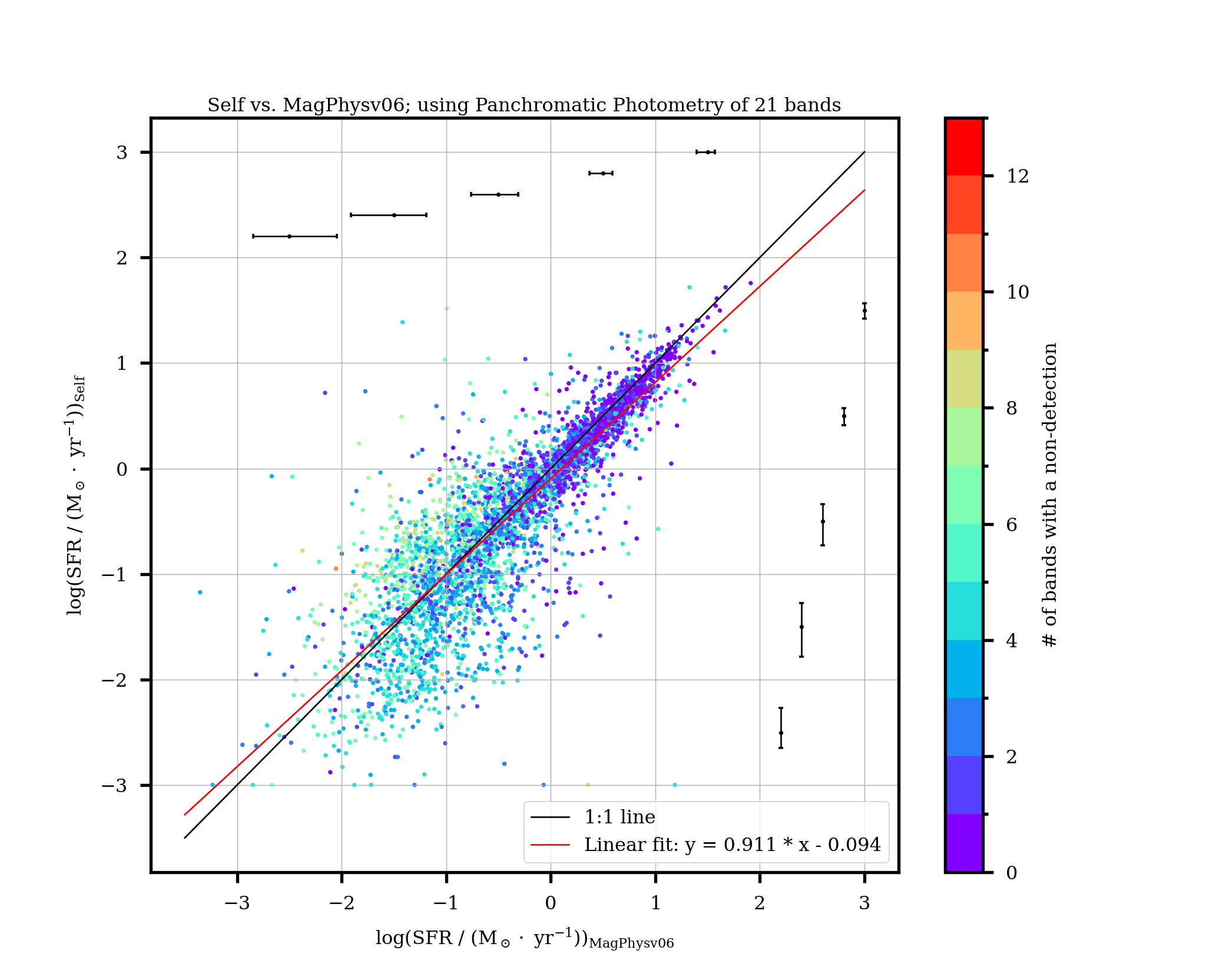}}}%
            \caption{A comparison of our results (``\texttt{Self}"), obtained using panchromatic photometry of 21 bands, with those from \texttt{MagPhysv06}: (a) logarithm of the stellar mass in units of M$_\odot$; (b) logarithm of the SFR in units of M$_\odot\cdot\rm{yr}^{-1}$. Each dot is a galaxy in the DESI-21Band sample. The colours dictate the number of bands with a non-detection, read from the colour bar. The error bars represent the mean error of the estimated quantities in the local bins defined by their positions and orientation. The black line is the 1:1 line, and the red line is the linear fit line.}%
            \label{fig:21band-magphys}%
        \end{figure}

\begin{figure}%
            \centering
            \subfloat[]{{\includegraphics[width=11cm]{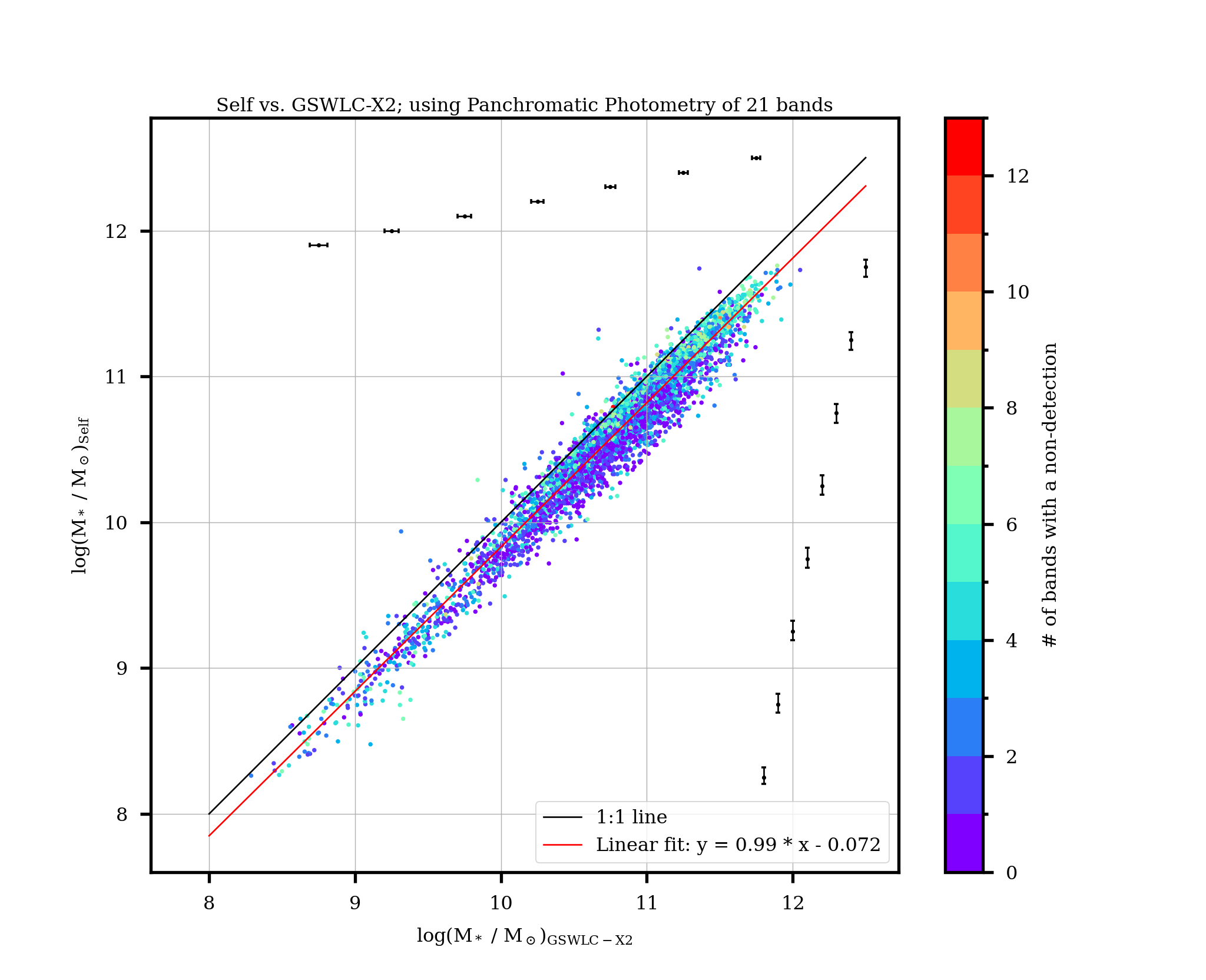}}}\\%
            \subfloat[]{{\includegraphics[width=11cm]{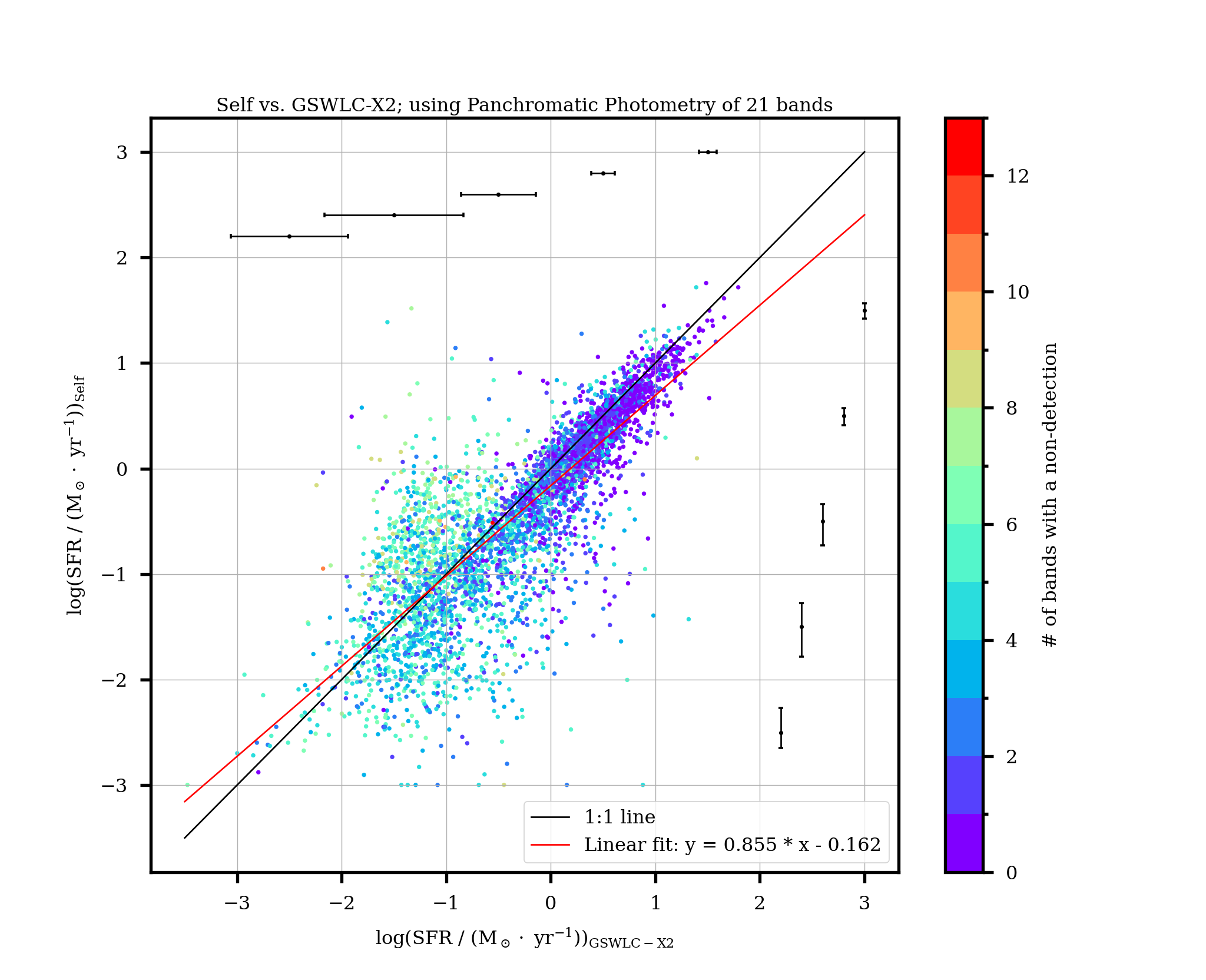}}}%
            \caption{Same as \figref{fig:21band-magphys}, except that here, the comparison is with \texttt{GSWLC-X2}.}%
            \label{fig:21band-gswlc}%
        \end{figure}

        We then compare with past studies, two of the most widely studied galaxy properties which are inferred from SED modelling: the stellar mass and the star formation rate (SFR). \figref{fig:21band-magphys} shows the comparison of the stellar masses and SFRs of the galaxies in the DESI-21Band sample, as inferred by us using MAGPHYS (referred to as ``$\texttt{Self}$"), with those obtained from the $\texttt{MagPhysv06}$ table. Our results agree with those from $\texttt{MagPhysv06}$ with no significant systematic difference in the stellar masses and the SFRs. \figref{fig:21band-gswlc} shows the same comparison with the $\texttt{GSWLC-X2}$ catalog (\cite{Salim16}, \cite{Salim18}). The stellar masses obtained by us are systematically less than those from $\texttt{GSWLC-X2}$ with a difference of $\sim0.2$ dex. The best fit line on the SFR comparison plot has a slope of $0.855$, which is mainly due to a large number of galaxies with 6 or more non-detection bands with $-1<\log(\rm{SFR / (M}_\odot\cdot\rm{yr}^{-1})_\texttt{Self}<0$. Comparing $\texttt{MagPhysv06}$ with $\texttt{GSWLC-X2}$ (plots not shown), we find similar trends in stellar masses and SFRs: difference of $\sim0.2$ dex in stellar masses and an SFR comparison best fit line with a slope of $0.833$. Since the agreement between our results and those from $\texttt{MagPhysv06}$ is remarkable, this difference might be because of the difference in the SPS modelling strategies implemented in MAGPHYS and those in CIGALE used by \cite{Salim18}. The scatter of the stellar masses about the best fit line is $\sim 0.15$ dex when compared with $\texttt{MagPhysv06}$ and $\sim 0.17$ when compared with $\texttt{GSWLC-X2}$. The scatter of the SFRs about the best fit line when compared with $\texttt{MagPhysv06}$ is $0.28$ dex for $\log(\rm{SFR / (M}_\odot\cdot\rm{yr}^{-1})_\texttt{MagPhysv06}>0$ and $0.50$ dex for $\log(\rm{SFR / (M}_\odot\cdot\rm{yr}^{-1})_\texttt{MagPhysv06}<0$, and when compared with $\texttt{GSWLC-X2}$, it is $0.35$ dex for $\log(\rm{SFR / (M}_\odot\cdot\rm{yr}^{-1})_\texttt{GSWLC-X2}>0$ and $0.58$ dex for $\log(\rm{SFR / (M}_\odot\cdot\rm{yr}^{-1})_\texttt{GSWLC-X2}<0$.
        
\begin{figure}%
            \centering
            \subfloat{{\includegraphics[width=17cm]{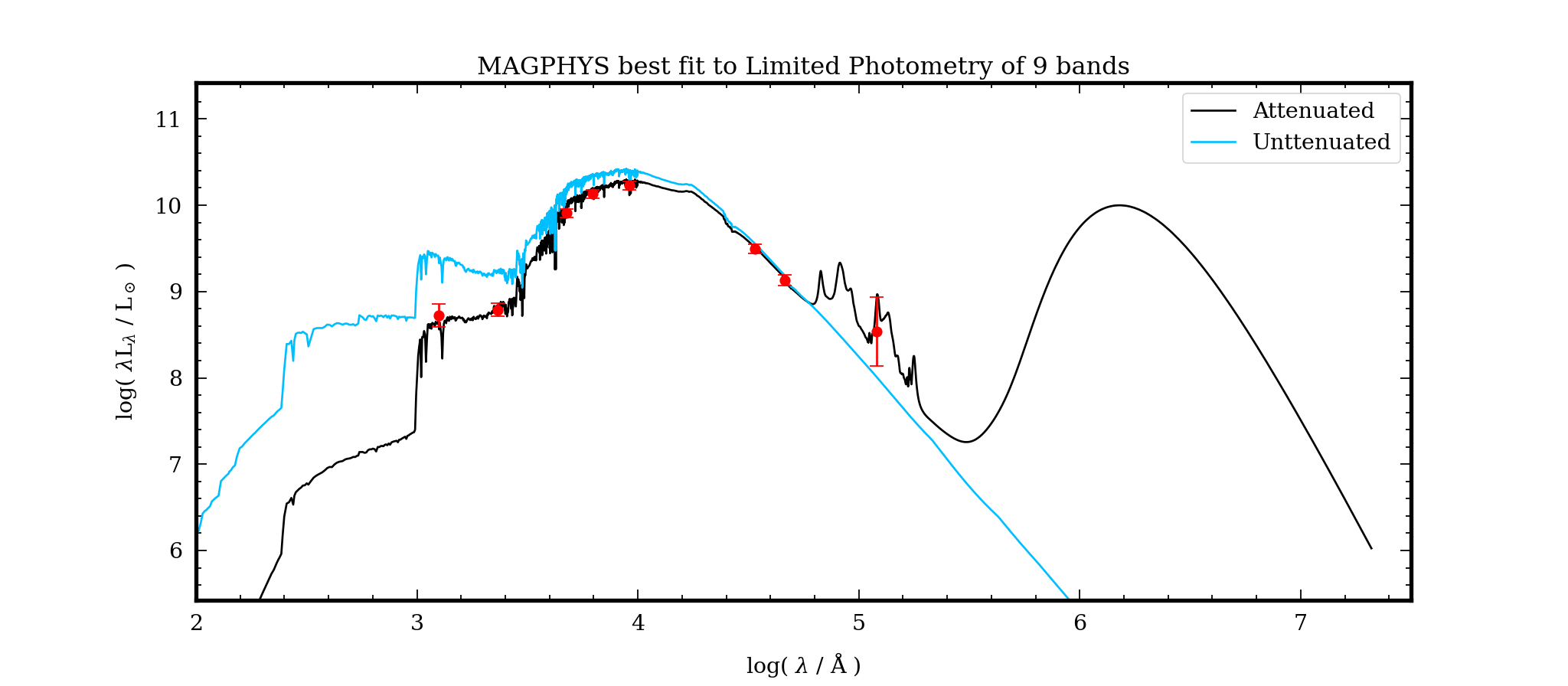}}}\\%
            \subfloat{{\includegraphics[width=17cm]{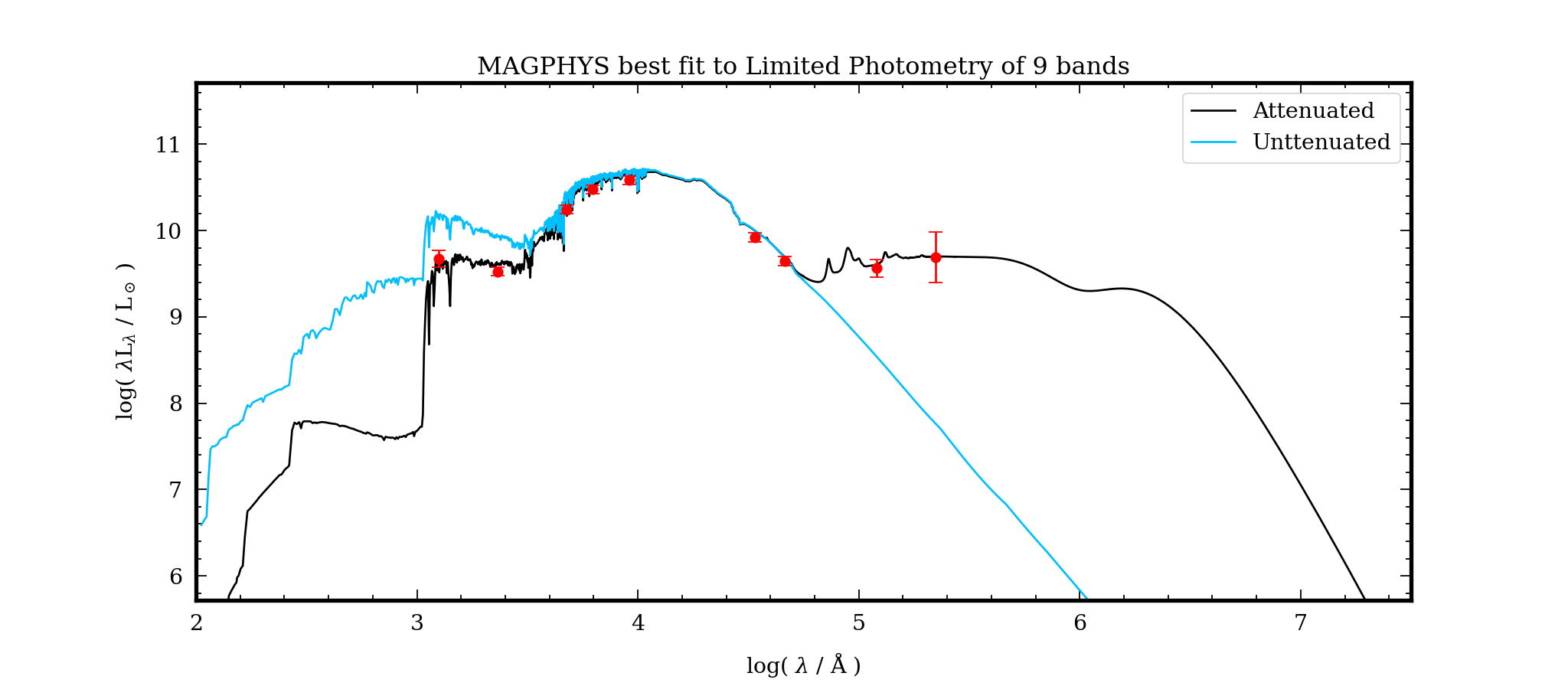}}}%
            \caption{Same as \figref{fig:21band-csp}, except that here, the results are obtained using limited photometry of 9 bands.}
            \label{fig:9band-csp}%
        \end{figure}

        After a successful demonstration of SPS modelling and SED fitting of the galaxies in the DESI-21Band sample using panchromatic photometry data, we did the same exercise with the same DESI-21Band sample, but with limited photometry of the 9 bands, namely the GALEX FUV and NUV bands, the $g, r$ and $z$ bands of the SDSS and the 4 NIR bands from WISE (W1-W4), which are common with the data in the DESI-GSWLC sample. The difference, however, is that the $g, r$ and $z$ bands in this new sample are taken from the DESI-21Band sample, which means that these are the $g, r$ and $z$ bands of the SDSS, but those in the DESI-GSWLC sample are taken from the DESI redshift summary value added catalog, which means they are the $g, r$ and $z$ bands of the DECaLS, BASS and MzLS Legacy Surveys. \figref{fig:9band-csp} shows the best fit CSPs of two randomly selected galaxies from the DESI-21Band sample using limited photometry of 9 bands. In this case also, the best fit CSPs (attenuated, black) are consistent with the observed SED (red circles) given the uncertainty in the observed fluxes (converted to luminosities) as represented by the errorbars and the agreement is similar to that obtained by the authors of MAGPHYS (see \figref{fig:sed_fitting}).
        
        \begin{figure}%
            \centering
            \subfloat[]{{\includegraphics[width=11cm]{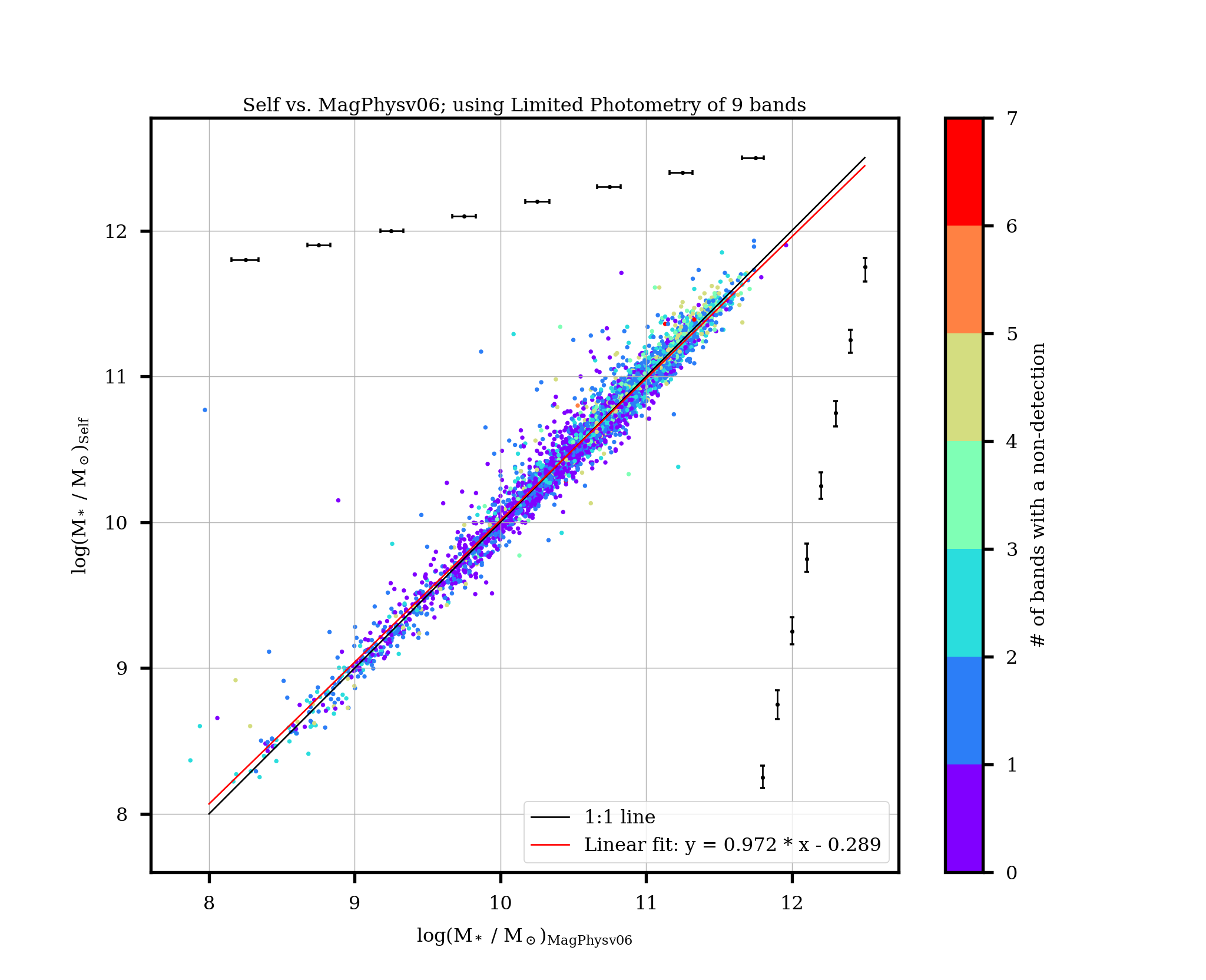}}}\\%
            \subfloat[]{{\includegraphics[width=11cm]{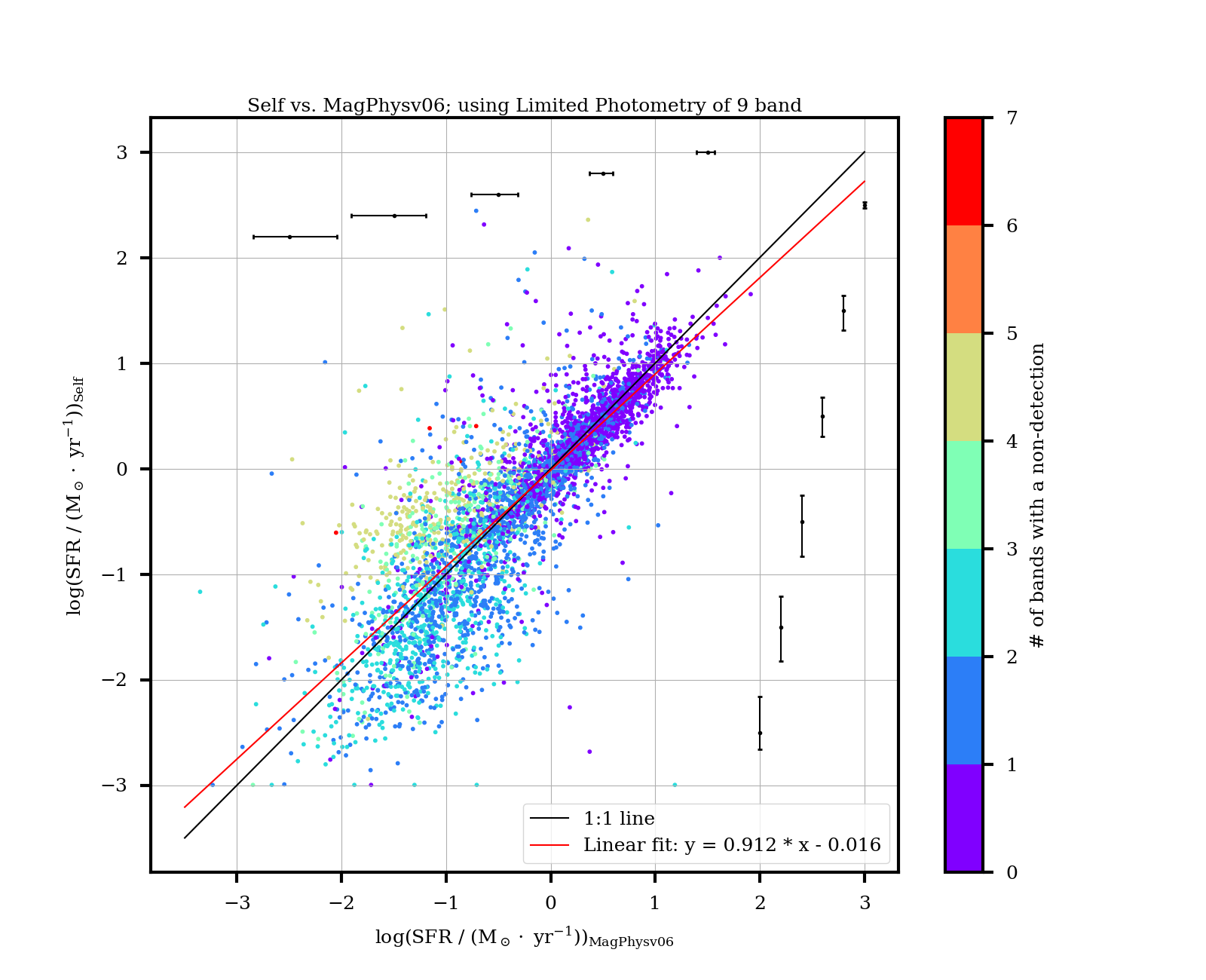}}}%
            \caption{Same as \figref{fig:21band-magphys}, except that here, the results are obtained using limited photometry of 9 bands.}%
            \label{fig:9band-magphys}%
        \end{figure}
        
        \begin{figure}%
            \centering
            \subfloat[]{{\includegraphics[width=11cm]{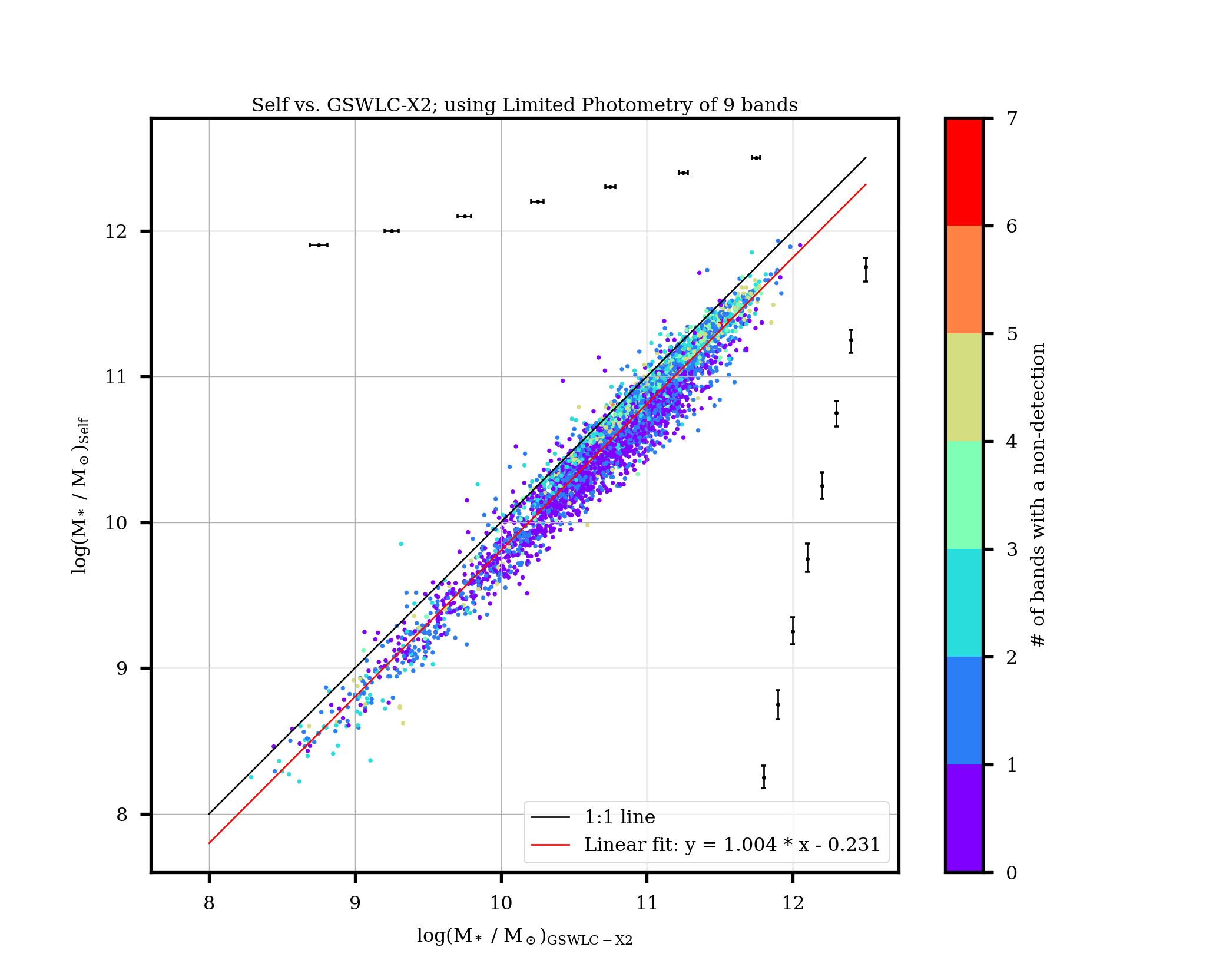}}}\\%
            \subfloat[]{{\includegraphics[width=11cm]{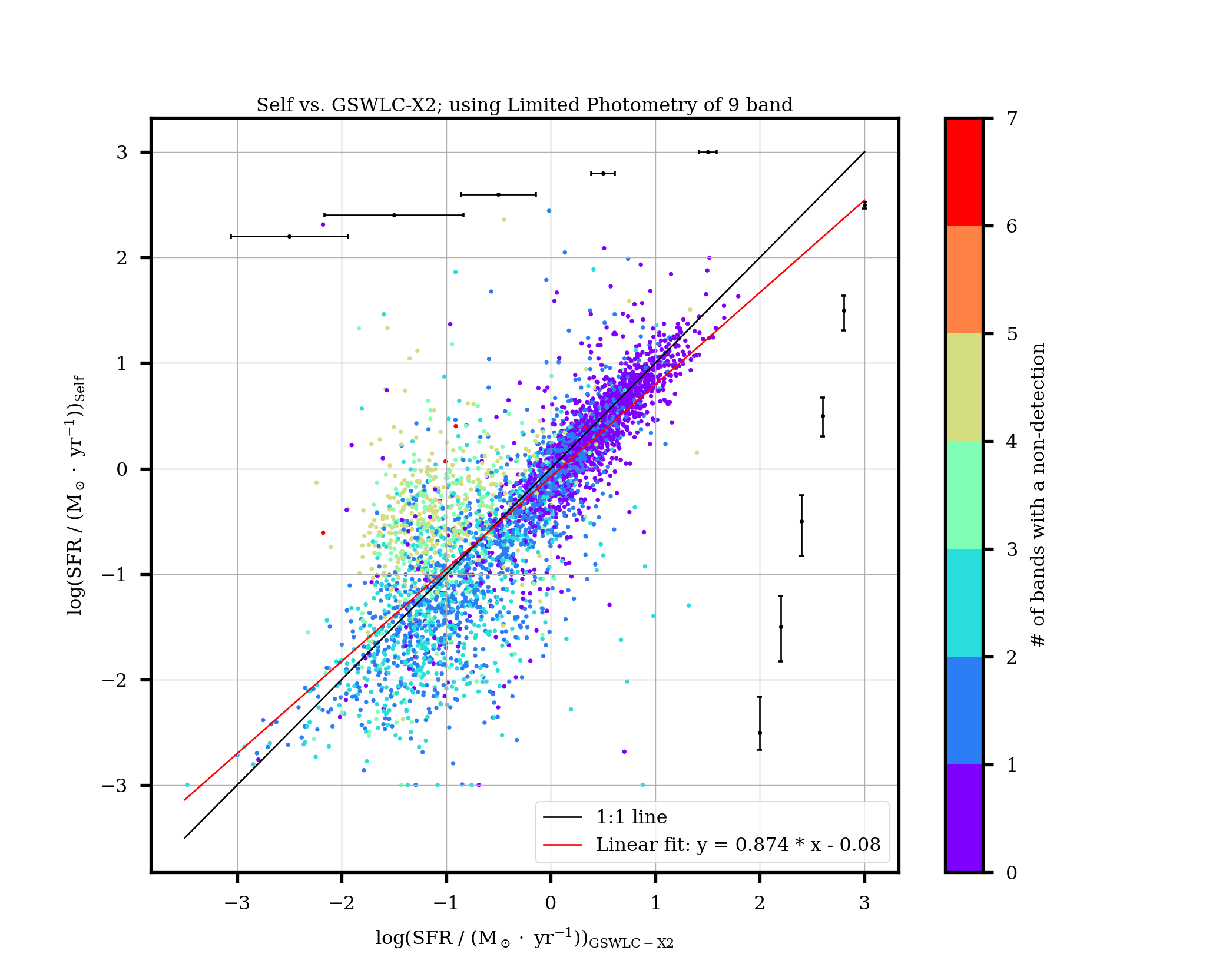}}}%
            \caption{Same as \figref{fig:9band-gswlc}, except that here, the comparison is with \texttt{GSWLC-X2}.}%
            \label{fig:9band-gswlc}%
        \end{figure}
        
        With the results obtained using limited photometry, we do the same comparison as mentioned above. \figref{fig:9band-magphys} shows the comparison of the stellar masses and SFRs of the galaxies in the DESI-21Band sample, as inferred by us using MAGPHYS using limited photometry, with those obtained from the $\texttt{MagPhysv06}$ table. Our results agree with those from $\texttt{MagPhysv06}$ with no significant systematic difference in the stellar masses and the SFRs. \figref{fig:9band-gswlc} shows the same comparison with the $\texttt{GSWLC-X2}$ catalog (\cite{Salim16}, \cite{Salim18}). The stellar masses obtained by us are systematically less than those from $\texttt{GSWLC-X2}$ with a difference of $\sim0.2$ dex. The best fit line on the SFR comparison plot has a slope of $0.874$, which is mainly due to a large number of galaxies with 4 or more non-detection bands with $-1<\log(\rm{SFR / (M}_\odot\cdot\rm{yr}^{-1})_\texttt{Self}<1$. Since this is a similar effect as discussed earlier, it might also be due to the differences in SPS modelling strategies used by MAGPHYS and \cite{Salim18}. The scatter of the stellar masses about the best fit line is $\sim 0.13$ dex when compared with $\texttt{MagPhysv06}$ and $\sim 0.14$ when compared with $\texttt{GSWLC-X2}$. The scatter of the SFRs about the best fit line when compared with $\texttt{MagPhysv06}$ is $0.32$ dex for $\log(\rm{SFR / (M}_\odot\cdot\rm{yr}^{-1})_\texttt{MagPhysv06}>0$ and $0.58$ dex for $\log(\rm{SFR / (M}_\odot\cdot\rm{yr}^{-1})_\texttt{MagPhysv06}<0$, and when compared with $\texttt{GSWLC-X2}$, it is $0.35$ dex for $\log(\rm{SFR / (M}_\odot\cdot\rm{yr}^{-1})_\texttt{GSWLC-X2}>0$ and $0.63$ dex for $\log(\rm{SFR / (M}_\odot\cdot\rm{yr}^{-1})_\texttt{GSWLC-X2}<0$.
        
        The above investigations suggest that the galaxy physical properties inferred from SPS modelling and SED fitting using limited photometric data of the previously mentioned 9 bands can be reliable if there are a maximum of 3 bands with non-detection. As we correctly expected, the UV bands (FUV and NUV) and the NIR bands (W1-W4) are found to be crucial for inferring the SFRs, the dust attenuation and emission. With that caution, one can draw correlations and other statistical inferences from such a catalog of galaxy physical parameters, even when full wavelength coverage of the SEDs is difficult to obtain because of practical limitations of the current generation telescopes and surveys.
        
    \section{Summary}\label{sec5.3}
        In this thesis, I first presented a bird's eye view of the broad field of galaxy formation and evolution studies, discussing in brief about various tools used in such studies and asserted that stellar population synthesis (SPS) modelling is one of such powerful tools. Details about SPS modelling, elaborating on the ingredients required for generating the SSPs and CSPs, were presented in \chref{ch2} and \chref{ch3}. The IMFs, stellar spectral libraries and the isochrone libraries combined give the SSP libraries and discussed in \secref{sec2.5}. SFH models, metallicity evolution models and dust attenuation and emission models combined with SSP libraries give the CSP libraries. Galaxy physical properties can be inferred by fitting their observed SEDs with the CSP libraries thus generated using statistical inference techniques like the Bayesian statistics. \partref{part2} of this thesis presents an overview of the Multi-wavelength Analysis of Galaxy PHYSical Properties (MAGPHYS) code (see \chref{ch4}) and the key findings from our SED fitting analysis of the DESI-21Band sample (see \chref{ch5}). \chref{ch4} presents a glimpse of numerous models available in the literature for different ingredients of SPS modelling and the implementation choices made in MAGPHYS. \chref{ch5} elaborates also the sample selection criteria for our investigations. We successfully inferred physical properties like stellar mass, SFR, dust attenuation and emission, etc. of the $4,391$ galaxies in the DESI-21Band sample. We first demonstrated this with panchromatic photometry of 21 bands from the PanchromaticPhotom DMU of the Galaxy and Mass Assembly (GAMA) \rnum{2} equatorial survey, and then with limited photometry of the 9 bands, namely the GALEX FUV and NUV bands, the g, r and z bands of the SDSS and the 4 NIR bands from WISE (W1-W4). The key finding from these investigations is that galaxy physical properties can be reliably inferred from SPS modelling and SED fitting using limited photometric data of the previously mentioned 9 bands if there are a maximum of 3 bands with non-detection, with the UV bands (FUV and NUV) and the NIR bands (W1-W4) playing a crucial role in inferring the SFRs and dust attenuation.
        
    \section{Future Work}\label{sec5.4}
        The immediate goal is to repeat the analysis on the DESI-GSWLC sample as a final verification of our approach. The next goal is to reliably infer the physical properties of the $2,97,402$ galaxies in the DESI-EDR sample with limited photometry of the 9 bands. We would do the same with galaxies in the Data Release 1 covering the first year of the DESI main survey data (through June 2022) that is expected to be publicly released in 2025, and hopefully, in the other future public data releases of the DESI main survey. Because of the enormously large number of galaxies expected to be observed in the DESI main survey and its cosmologically large volume coverage, such an analysis might prove to be the next milestone in the field of galaxy formation and evolution.

%% file: ref.bib
@article{Bait17,
    author = {{Bait}, Omkar and {Barway}, Sudhanshu and {Wadadekar}, Yogesh},
    title = "{On the interdependence of galaxy morphology, star formation and environment in massive galaxies in the nearby Universe}",
    journal = {Monthly Notices of the Royal Astronomical Society},
    volume = {471},
    number = {3},
    pages = {2687-2702},
    year = {2017},
    month = {07},
    issn = {0035-8711},
    doi = {10.1093/mnras/stx1688},
    url = {https://doi.org/10.1093/mnras/stx1688},
    eprint = {https://academic.oup.com/mnras/article-pdf/471/3/2687/19496858/stx1688.pdf},
}

@article{Conroy13,
       author = {{Conroy}, Charlie},
        title = "{Modeling the Panchromatic Spectral Energy Distributions of Galaxies}",
      journal = {Annual Review of Astronomy and Astrophysics},
         year = 2013,
        month = {08},
       volume = {51},
       number = {1},
        pages = {393-455},
          doi = {10.1146/annurev-astro-082812-141017},
archivePrefix = {arXiv},
       eprint = {1301.7095},
 primaryClass = {astro-ph.CO},
       adsurl = {https://ui.adsabs.harvard.edu/abs/2013ARAandA..51..393C}
}

@article{daCunha08,
       author = {{da Cunha}, Elisabete and {Charlot}, St{\'e}phane and {Elbaz}, David},
        title = "{A simple model to interpret the ultraviolet, optical and infrared emission from galaxies}",
      journal = {Monthly Notices of the Royal Astronomical Society},
         year = 2008,
        month = {08},
       volume = {388},
       number = {4},
        pages = {1595-1617},
          doi = {10.1111/j.1365-2966.2008.13535.x},
archivePrefix = {arXiv},
       eprint = {0806.1020},
 primaryClass = {astro-ph},
       adsurl = {https://ui.adsabs.harvard.edu/abs/2008MNRAS.388.1595D}
}

@article{Chabrier03,
       author = {{Chabrier}, Gilles},
        title = "{Galactic Stellar and Substellar Initial Mass Function}",
      journal = {Publications of the Astronomical Society of the Pacific},
         year = 2003,
        month = {07},
       volume = {115},
       number = {809},
        pages = {763-795},
          doi = {10.1086/376392},
archivePrefix = {arXiv},
       eprint = {astro-ph/0304382},
 primaryClass = {astro-ph},
       adsurl = {https://ui.adsabs.harvard.edu/abs/2003PASP..115..763C}
}

@article{Salpeter55,
       author = {{Salpeter}, Edwin E.},
        title = "{The Luminosity Function and Stellar Evolution.}",
      journal = {The Astrophysical Journal},
         year = 1955,
        month = {01},
       volume = {121},
        pages = {161},
          doi = {10.1086/145971},
       adsurl = {https://ui.adsabs.harvard.edu/abs/1955ApJ...121..161S}
}

@article{MillerScalo79,
       author = {{Miller}, G.~E. and {Scalo}, J.~M.},
        title = "{The Initial Mass Function and Stellar Birthrate in the Solar Neighborhood}",
      journal = {The Astrophysical Journal Supplement Series},
         year = 1979,
        month = {11},
       volume = {41},
        pages = {513},
          doi = {10.1086/190629},
       adsurl = {https://ui.adsabs.harvard.edu/abs/1979ApJS...41..513M}
}

@article{Kroupa01,
       author = {{Kroupa}, Pavel},
        title = "{On the variation of the initial mass function}",
      journal = {Monthly Notices of the Royal Astronomical Society},
         year = 2001,
        month = {04},
       volume = {322},
       number = {2},
        pages = {231-246},
          doi = {10.1046/j.1365-8711.2001.04022.x},
archivePrefix = {arXiv},
       eprint = {astro-ph/0009005},
 primaryClass = {astro-ph},
       adsurl = {https://ui.adsabs.harvard.edu/abs/2001MNRAS.322..231K}
}

@article{Jacoby84,
       author = {{Jacoby}, G.~H. and {Hunter}, D.~A. and {Christian}, C.~A.},
        title = "{A library of stellar spectra.}",
      journal = {The Astrophysical Journal Supplement Series},
         year = 1984,
        month = {10},
       volume = {56},
        pages = {257-281},
          doi = {10.1086/190983},
       adsurl = {https://ui.adsabs.harvard.edu/abs/1984ApJS...56..257J}
}

@article{Cardelli89,
       author = {{Cardelli}, Jason A. and {Clayton}, Geoffrey C. and {Mathis}, John S.},
        title = "{The Relationship between Infrared, Optical, and Ultraviolet Extinction}",
      journal = {The Astrophysical Journal},
         year = 1989,
        month = {10},
       volume = {345},
        pages = {245},
          doi = {10.1086/167900},
       adsurl = {https://ui.adsabs.harvard.edu/abs/1989ApJ...345..245C}
}

@article{Fitzpatrick99,
doi = {10.1086/316293},
url = {https://dx.doi.org/10.1086/316293},
year = {1999},
month = {01},
publisher = {The University of Chicago Press},
volume = {111},
number = {755},
pages = {63},
author = {{Fitzpatrick}, Edward L.},
title = {Correcting for the Effects of Interstellar Extinction},
journal = {Publications of the Astronomical Society of the Pacific}
}

@article{Seaton79,
    author = {{Seaton}, M. J.},
    title = "{Interstellar extinction in the UV}",
    journal = {Monthly Notices of the Royal Astronomical Society},
    volume = {187},
    number = {1},
    pages = {73P-76P},
    year = {1979},
    month = {05},
    issn = {0035-8711},
    doi = {10.1093/mnras/187.1.73P},
    url = {https://doi.org/10.1093/mnras/187.1.73P},
    eprint = {https://academic.oup.com/mnras/article-pdf/187/1/73P/3786563/mnras187-073P.pdf},
}

@article{BaSTI1,
       author = {{Hidalgo}, Sebastian L. and {Pietrinferni}, Adriano and {Cassisi}, Santi and {Salaris}, Maurizio and {Mucciarelli}, Alessio and {Savino}, Alessandro and {Aparicio}, Antonio and {Silva Aguirre}, Victor and {Verma}, Kuldeep},
        title = "{The Updated BaSTI Stellar Evolution Models and Isochrones. I. Solar-scaled Calculations}",
      journal = {The Astrophysical Journal},
         year = 2018,
        month = {04},
       volume = {856},
       number = {2},
          eid = {125},
        pages = {125},
          doi = {10.3847/1538-4357/aab158},
archivePrefix = {arXiv},
       eprint = {1802.07319},
 primaryClass = {astro-ph.GA},
       adsurl = {https://ui.adsabs.harvard.edu/abs/2018ApJ...856..125H}
}

@article{BaSTI2,
doi = {10.3847/1538-4357/abd4d5},
url = {https://dx.doi.org/10.3847/1538-4357/abd4d5},
year = {2021},
month = {02},
publisher = {The American Astronomical Society},
volume = {908},
number = {1},
pages = {102},
author = {Adriano Pietrinferni and Sebastian Hidalgo and Santi Cassisi and Maurizio Salaris and Alessandro Savino and Alessio Mucciarelli and Kuldeep Verma and Victor Silva Aguirre and Antonio Aparicio and Jason W. Ferguson},
title = {Updated BaSTI Stellar Evolution Models and Isochrones. II. {\ensuremath{\alpha}}-enhanced Calculations},
journal = {The Astrophysical Journal}
}

@article{BaSTI3,
    author = {Salaris, Maurizio and Cassisi, Santi and Pietrinferni, Adriano and Hidalgo, Sebastian},
    title = {The updated basti stellar evolution models and isochrones – III. White dwarfs},
    journal = {Monthly Notices of the Royal Astronomical Society},
    volume = {509},
    number = {4},
    pages = {5197-5208},
    year = {2021},
    month = {11},
    issn = {0035-8711},
    doi = {10.1093/mnras/stab3359},
    url = {https://doi.org/10.1093/mnras/stab3359},
    eprint = {https://academic.oup.com/mnras/article-pdf/509/4/5197/41721761/stab3359.pdf},
}

@article{BaSTI4,
    author = {Pietrinferni, Adriano and Salaris, Maurizio and Cassisi, Santi and Savino, Alessandro and Mucciarelli, Alessio and Hyder, David and Hidalgo, Sebastian},
    title = {The updated BaSTI stellar evolution models and isochrones – IV. {\ensuremath{\alpha}}-Depleted calculations},
    journal = {Monthly Notices of the Royal Astronomical Society},
    volume = {527},
    number = {2},
    pages = {2065-2070},
    year = {2023},
    month = {10},
    issn = {0035-8711},
    doi = {10.1093/mnras/stad3267},
    url = {https://doi.org/10.1093/mnras/stad3267},
    eprint = {https://academic.oup.com/mnras/article-pdf/527/2/2065/53273641/stad3267.pdf},
}

@article{BC03,
       author = {{Bruzual}, G. and {Charlot}, S.},
        title = {Stellar population synthesis at the resolution of 2003},
      journal = {Monthly Notices of the Royal Astronomical Society},
         year = 2003,
        month = {10},
       volume = {344},
       number = {4},
        pages = {1000-1028},
          doi = {10.1046/j.1365-8711.2003.06897.x},
archivePrefix = {arXiv},
       eprint = {astro-ph/0309134},
 primaryClass = {astro-ph},
       adsurl = {https://ui.adsabs.harvard.edu/abs/2003MNRAS.344.1000B}
}

@article{Boquien19,
       author = {{Boquien}, M. and {Burgarella}, D. and {Roehlly}, Y. and {Buat}, V. and {Ciesla}, L. and {Corre}, D. and {Inoue}, A.~K. and {Salas}, H.},
        title = {CIGALE: a python Code Investigating GALaxy Emission},
      journal = {Astronomy and Astrophysics},
         year = 2019,
        month = {02},
       volume = {622},
          eid = {A103},
        pages = {A103},
          doi = {10.1051/0004-6361/201834156},
archivePrefix = {arXiv},
       eprint = {1811.03094},
 primaryClass = {astro-ph.GA},
       adsurl = {https://ui.adsabs.harvard.edu/abs/2019AandA...622A.103B}
}

@article{Dwek97,
doi = {10.1086/303568},
url = {https://dx.doi.org/10.1086/303568},
year = {1997},
month = {02},
publisher = {},
volume = {475},
number = {2},
pages = {565},
author = {E. Dwek and R. G. Arendt and D. J. Fixsen and T. J. Sodroski and N. Odegard and J. L. Weiland and W. T. Reach and M. G. Hauser and T. Kelsall and S. H. Moseley and R. F. Silverberg and R. A. Shafer and J. Ballester and D. Bazell and R. Isaacman},
title = {Detection and Characterization of Cold Interstellar Dust and Polycyclic Aromatic Hydrocarbon Emission, from COBE Observations},
journal = {The Astrophysical Journal},
}

@article{Flagey06,
       author = {{Flagey}, N. and {Boulanger}, F. and {Verstraete}, L. and {Miville Desch{\^e}nes}, M.~A. and {Noriega Crespo}, A. and {Reach}, W.~T.},
        title = {Spitzer/IRAC and ISOCAM/CVF insights on the origin of the near to mid-IR Galactic diffuse emission},
      journal = {Astronomy and Astrophysics},
         year = 2006,
        month = {07},
       volume = {453},
       number = {3},
        pages = {969-978},
          doi = {10.1051/0004-6361:20053949},
archivePrefix = {arXiv},
       eprint = {astro-ph/0604238},
 primaryClass = {astro-ph},
       adsurl = {https://ui.adsabs.harvard.edu/abs/2006AandA...453..969F}
}

@article{Saha1920,
author = {{Saha}, Megh Nad},
title = {LIII. Ionization in the solar chromosphere },
journal = {The London, Edinburgh, and Dublin Philosophical Magazine and Journal of Science},
volume = {40},
number = {238},
pages = {472--488},
year = {1920},
publisher = {Taylor \and Francis},
doi = {10.1080/14786441008636148},
URL = {https://doi.org/10.1080/14786441008636148},
eprint = {https://doi.org/10.1080/14786441008636148}
}

@article{Baumann10,
       author = {{Baumann}, P. and {Ram{\'\i}rez}, I. and {Mel{\'e}ndez}, J. and {Asplund}, M. and {Lind}, K.},
        title = "{Lithium depletion in solar-like stars: no planet connection}",
      journal = {Astronomy and Astrophysics},
         year = 2010,
        month = {09},
       volume = {519},
          eid = {A87},
        pages = {A87},
          doi = {10.1051/0004-6361/201015137},
archivePrefix = {arXiv},
       eprint = {1008.0575},
 primaryClass = {astro-ph.SR},
       adsurl = {https://ui.adsabs.harvard.edu/abs/2010AandA...519A..87B}
}

@article{Hertzsprung1911,
  title={On the Use of Photographic Effective Wavelengths for the Determination of Color Equivalents},
  author={Hertzsprung, E},
  journal={Publications of the Astrophysical Observatory in Potsdam},
  volume={1},
  pages={22--63},
  year={1911}
}

@article{Russell1913,
  title={"Giant" and "dwarf" stars},
  author={Russell, Henry Norris},
  journal={The Observatory},
  volume={36},
  pages={324--329},
  year={1913}
}

@article{Calzetti07,
       author = {{Calzetti}, D. and {Kennicutt}, R.~C. and {Engelbracht}, C.~W. and {Leitherer}, C. and {Draine}, B.~T. and {Kewley}, L. and {Moustakas}, J. and {Sosey}, M. and {Dale}, D.~A. and {Gordon}, K.~D. and {Helou}, G.~X. and {Hollenbach}, D.~J. and {Armus}, L. and {Bendo}, G. and {Bot}, C. and {Buckalew}, B. and {Jarrett}, T. and {Li}, A. and {Meyer}, M. and {Murphy}, E.~J. and {Prescott}, M. and {Regan}, M.~W. and {Rieke}, G.~H. and {Roussel}, H. and {Sheth}, K. and {Smith}, J.~D.~T. and {Thornley}, M.~D. and {Walter}, F.},
        title = {The Calibration of Mid-Infrared Star Formation Rate Indicators},
      journal = {The Astrophysical Journal},
         year = 2007,
        month = {09},
       volume = {666},
       number = {2},
        pages = {870-895},
          doi = {10.1086/520082},
archivePrefix = {arXiv},
       eprint = {0705.3377},
 primaryClass = {astro-ph},
       adsurl = {https://ui.adsabs.harvard.edu/abs/2007ApJ...666..870C}
}

@article{Kennicutt07,
doi = {10.1086/522300},
url = {https://dx.doi.org/10.1086/522300},
year = {2007},
month = {12},
publisher = {},
volume = {671},
number = {1},
pages = {333},
author = {Robert C. Kennicutt, Jr. and Daniela Calzetti and Fabian Walter and George Helou and David J. Hollenbach and Lee Armus and George Bendo and Daniel A. Dale and Bruce T. Draine and Charles W. Engelbracht and Karl D. Gordon and Moire K. M. Prescott and Michael W. Regan and Michele D. Thornley and Caroline Bot and Elias Brinks and Erwin de Blok and Dulia de Mello and Martin Meyer and John Moustakas and Eric J. Murphy and Kartik Sheth and J. D. T. Smith},
title = {Star Formation in NGC 5194 (M51a). II. The Spatially Resolved Star Formation Law},
journal = {The Astrophysical Journal}
}

@article{B2FH,
  title = {Synthesis of the Elements in Stars},
  author = {Burbidge, E. Margaret and Burbidge, G. R. and Fowler, William A. and Hoyle, F.},
  journal = {Reviews of Modern Physics},
  volume = {29},
  issue = {4},
  pages = {547--650},
  numpages = {0},
  year = 1957,
  month = {10},
  publisher = {American Physical Society},
  doi = {10.1103/RevModPhys.29.547},
  url = {https://link.aps.org/doi/10.1103/RevModPhys.29.547}
}

@article{Asano13,
    author = {Asano, Ryosuke S. and Takeuchi, Tsutomu T. and Hirashita, Hiroyuki and Inoue, Akio K.},
    title = {Dust formation history of galaxies: A critical role of metallicity* for the dust mass growth by accreting materials in the interstellar medium"},
    journal = {Earth, Planets and Space},
    year = 2013,
    month = {03},
    volume = {65},
    issue = {3},
    pages = {213-222},
    url = {https://doi.org/10.5047/eps.2012.04.014},
    doi = {10.5047/eps.2012.04.014}
}

@article{Yates12,
       author = {{Yates}, Robert M. and {Kauffmann}, Guinevere and {Guo}, Qi},
        title = "{The relation between metallicity, stellar mass and star formation in galaxies: an analysis of observational and model data}",
      journal = {Monthly Notices of the Royal Astronomical Society},
         year = 2012,
        month = {05},
       volume = {422},
       number = {1},
        pages = {215-231},
          doi = {10.1111/j.1365-2966.2012.20595.x},
archivePrefix = {arXiv},
       eprint = {1107.3145},
 primaryClass = {astro-ph.CO},
       adsurl = {https://ui.adsabs.harvard.edu/abs/2012MNRAS.422..215Y}
}

@article{Pacifici23,
doi = {10.3847/1538-4357/acacff},
url = {https://dx.doi.org/10.3847/1538-4357/acacff},
year = {2023},
month = {02},
publisher = {The American Astronomical Society},
volume = {944},
number = {2},
pages = {141},
author = {Camilla Pacifici and Kartheik G. Iyer and Bahram Mobasher and Elisabete da Cunha and Viviana Acquaviva and Denis Burgarella and Gabriela Calistro Rivera and Adam C. Carnall and Yu-Yen Chang and Nima Chartab and Kevin C. Cooke and Ciaran Fairhurst and Jeyhan Kartaltepe and Joel Leja and Katarzyna Małek and Brett Salmon and Marianna Torelli and Alba Vidal-García and Médéric Boquien and Gabriel G. Brammer and Michael J. I. Brown and Peter L. Capak and Jacopo Chevallard and Chiara Circosta and Darren Croton and Iary Davidzon and Mark Dickinson and Kenneth J. Duncan and Sandra M. Faber and Harry C. Ferguson and Adriano Fontana and Yicheng Guo and Boris Haeussler and Shoubaneh Hemmati and Marziye Jafariyazani and Susan A. Kassin and Rebecca L. Larson and Bomee Lee and Kameswara Bharadwaj Mantha and Francesca Marchi and Hooshang Nayyeri and Jeffrey A. Newman and Viraj Pandya and Janine Pforr and Naveen Reddy and Ryan Sanders and Ekta Shah and Abtin Shahidi and Matthew L. Stevans and Dian Puspita Triani and Krystal D. Tyler and Brittany N. Vanderhoof and Alexander de la Vega and Weichen Wang and Madalyn E. Weston},
title = {The Art of Measuring Physical Parameters in Galaxies: A Critical Assessment of Spectral Energy Distribution Fitting Techniques},
journal = {The Astrophysical Journal}
}

@article{BAGPIPES,
    author = {Carnall, A C and McLure, R J and Dunlop, J S and Davé, R},
    title = "{Inferring the star formation histories of massive quiescent galaxies with bagpipes: evidence for multiple quenching mechanisms}",
    journal = {Monthly Notices of the Royal Astronomical Society},
    volume = {480},
    number = {4},
    pages = {4379-4401},
    year = {2018},
    month = {08},
    issn = {0035-8711},
    doi = {10.1093/mnras/sty2169},
    url = {https://doi.org/10.1093/mnras/sty2169},
    eprint = {https://academic.oup.com/mnras/article-pdf/480/4/4379/25539546/sty2169.pdf},
}

@article{Prospector1,
doi = {10.3847/1538-4357/aa5ffe},
url = {https://dx.doi.org/10.3847/1538-4357/aa5ffe},
year = {2017},
month = {03},
publisher = {The American Astronomical Society},
volume = {837},
number = {2},
pages = {170},
author = {Joel Leja and Benjamin D. Johnson and Charlie Conroy and Pieter G. van Dokkum and Nell Byler},
title = {Deriving Physical Properties from Broadband Photometry with Prospector: Description of the Model and a Demonstration of its Accuracy Using 129 Galaxies in the Local Universe},
journal = {The Astrophysical Journal}
}

@article{Prospector2,
doi = {10.3847/1538-4365/abef67},
url = {https://dx.doi.org/10.3847/1538-4365/abef67},
year = {2021},
month = {05},
publisher = {The American Astronomical Society},
volume = {254},
number = {2},
pages = {22},
author = {Benjamin D. Johnson and Joel Leja and Charlie Conroy and Joshua S. Speagle},
title = {Stellar Population Inference with Prospector},
journal = {The Astrophysical Journal Supplement Series}
}

@article{GRASIL,
doi = {10.1086/306476},
url = {https://dx.doi.org/10.1086/306476},
year = {1998},
month = {12},
publisher = {},
volume = {509},
number = {1},
pages = {103},
author = {Laura Silva and Gian Luigi Granato and Alessandro Bressan and Luigi Danese},
title = {Modeling the Effects of Dust on Galactic Spectral Energy Distributions from the Ultraviolet to the Millimeter Band},
journal = {The Astrophysical Journal}}

@article{Yan19,
       author = {{Yan}, Renbin and {Chen}, Yanping and {Lazarz}, Daniel and {Bizyaev}, Dmitry and {Maraston}, Claudia and {Stringfellow}, Guy S. and {McCarthy}, Kyle and {Meneses-Goytia}, Sofia and {Law}, David R. and {Thomas}, Daniel and {Falcon Barroso}, Jesus and {S{\'a}nchez-Gallego}, Jos{\'e} R. and {Schlafly}, Edward and {Zheng}, Zheng and {Argudo-Fern{\'a}ndez}, Maria and {Beaton}, Rachael L. and {Beers}, Timothy C. and {Bershady}, Matthew and {Blanton}, Michael R. and {Brownstein}, Joel and {Bundy}, Kevin and {Chambers}, Kenneth C. and {Cherinka}, Brian and {De Lee}, Nathan and {Drory}, Niv and {Galbany}, Llu{\'\i}s and {Holtzman}, Jon and {Imig}, Julie and {Kaiser}, Nick and {Kinemuchi}, Karen and {Liu}, Chao and {Luo}, A. -Li and {Magnier}, Eugene and {Majewski}, Steven and {Nair}, Preethi and {Oravetz}, Audrey and {Oravetz}, Daniel and {Pan}, Kaike and {Sobeck}, Jennifer and {Stassun}, Keivan and {Talbot}, Michael and {Tremonti}, Christy and {Waters}, Christopher and {Weijmans}, Anne-Marie and {Wilhelm}, Ronald and {Zasowski}, Gail and {Zhao}, Gang and {Zhao}, Yong-Heng},
        title = "{SDSS-IV MaStar: A Large and Comprehensive Empirical Stellar Spectral Library{\textemdash}First Release}",
      journal = {The Astrophysical Journal},
         year = 2019,
        month = {10},
       volume = {883},
       number = {2},
          eid = {175},
        pages = {175},
          doi = {10.3847/1538-4357/ab3ebc},
archivePrefix = {arXiv},
       eprint = {1812.02745},
 primaryClass = {astro-ph.IM},
       adsurl = {https://ui.adsabs.harvard.edu/abs/2019ApJ...883..175Y}
}

@article{Maraston05,
       author = {{Maraston}, Claudia},
        title = "{Evolutionary population synthesis: models, analysis of the ingredients and application to high-z galaxies}",
      journal = {Monthly Notices of the Royal Astronomical Society},
         year = 2005,
        month = {09},
       volume = {362},
       number = {3},
        pages = {799-825},
          doi = {10.1111/j.1365-2966.2005.09270.x},
archivePrefix = {arXiv},
       eprint = {astro-ph/0410207},
 primaryClass = {astro-ph},
       adsurl = {https://ui.adsabs.harvard.edu/abs/2005MNRAS.362..799M}
}

@article{Buat08,
       author = {{Buat}, V. and {Boissier}, S. and {Burgarella}, D. and {Takeuchi}, T.~T. and {Le Floc'h}, E. and {Marcillac}, D. and {Huang}, J. and {Nagashima}, M. and {Enoki}, M.},
        title = "{Star formation history of galaxies from z = 0 to z = 0.7. A backward approach to the evolution of star-forming galaxies}",
      journal = {Astronomy and Astrophysics},
         year = 2008,
        month = {05},
       volume = {483},
       number = {1},
        pages = {107-119},
          doi = {10.1051/0004-6361:20078263},
archivePrefix = {arXiv},
       eprint = {0803.0414},
 primaryClass = {astro-ph},
       adsurl = {https://ui.adsabs.harvard.edu/abs/2008AandA...483..107B}
}

@article{Buat18,
       author = {{Buat}, V. and {Boquien}, M. and {Ma{\l}ek}, K. and {Corre}, D. and {Salas}, H. and {Roehlly}, Y. and {Shirley}, R. and {Efstathiou}, A.},
        title = "{Dust attenuation and H{\ensuremath{\alpha}} emission in a sample of galaxies observed with Herschel at 0.6 < z < 1.6}",
      journal = {Astronomy and Astrophysics},
         year = 2018,
        month = {11},
       volume = {619},
          eid = {A135},
        pages = {A135},
          doi = {10.1051/0004-6361/201833841},
archivePrefix = {arXiv},
       eprint = {1809.00161},
 primaryClass = {astro-ph.GA},
       adsurl = {https://ui.adsabs.harvard.edu/abs/2018AandA...619A.135B}
}

@article{Marigo07,
       author = {{Marigo}, P. and {Girardi}, L.},
        title = "{Evolution of asymptotic giant branch stars. I. Updated synthetic TP-AGB models and their basic calibration}",
      journal = {Astronomy and Astrophysics},
         year = 2007,
        month = {07},
       volume = {469},
       number = {1},
        pages = {239-263},
          doi = {10.1051/0004-6361:20066772},
archivePrefix = {arXiv},
       eprint = {astro-ph/0703139},
 primaryClass = {astro-ph},
       adsurl = {https://ui.adsabs.harvard.edu/abs/2007AandA...469..239M}
}

@article{CF00,
       author = {{Charlot}, St{\'e}phane and {Fall}, S. Michael},
        title = "{A Simple Model for the Absorption of Starlight by Dust in Galaxies}",
      journal = {The Astrophysical Journal},
         year = 2000,
        month = {08},
       volume = {539},
       number = {2},
        pages = {718-731},
          doi = {10.1086/309250},
archivePrefix = {arXiv},
       eprint = {astro-ph/0003128},
 primaryClass = {astro-ph},
       adsurl = {https://ui.adsabs.harvard.edu/abs/2000ApJ...539..718C}
}

@article{Draine07,
       author = {{Draine}, B.~T. and {Li}, Aigen},
        title = "{Infrared Emission from Interstellar Dust. IV. The Silicate-Graphite-PAH Model in the Post-Spitzer Era}",
      journal = {The Astrophysical Journal},
         year = 2007,
        month = {03},
       volume = {657},
       number = {2},
        pages = {810-837},
          doi = {10.1086/511055},
archivePrefix = {arXiv},
       eprint = {astro-ph/0608003},
 primaryClass = {astro-ph},
       adsurl = {https://ui.adsabs.harvard.edu/abs/2007ApJ...657..810D}
}

@article{Draine14,
       author = {{Draine}, B.~T. and {Aniano}, G. and {Krause}, Oliver and {Groves}, Brent and {Sandstrom}, Karin and {Braun}, Robert and {Leroy}, Adam and {Klaas}, Ulrich and {Linz}, Hendrik and {Rix}, Hans-Walter and {Schinnerer}, Eva and {Schmiedeke}, Anika and {Walter}, Fabian},
        title = "{Andromeda's Dust}",
      journal = {The Astrophysical Journal},
         year = 2014,
        month = {01},
       volume = {780},
       number = {2},
          eid = {172},
        pages = {172},
          doi = {10.1088/0004-637X/780/2/172},
archivePrefix = {arXiv},
       eprint = {1306.2304},
 primaryClass = {astro-ph.CO},
       adsurl = {https://ui.adsabs.harvard.edu/abs/2014ApJ...780..172D}
}

@article{Dale14,
       author = {{Dale}, Daniel A. and {Helou}, George and {Magdis}, Georgios E. and {Armus}, Lee and {D{\'\i}az-Santos}, Tanio and {Shi}, Yong},
        title = "{A Two-parameter Model for the Infrared/Submillimeter/Radio Spectral Energy Distributions of Galaxies and Active Galactic Nuclei}",
      journal = {The Astrophysical Journal},
         year = 2014,
        month = {03},
       volume = {784},
       number = {1},
          eid = {83},
        pages = {83},
          doi = {10.1088/0004-637X/784/1/83},
archivePrefix = {arXiv},
       eprint = {1402.1495},
 primaryClass = {astro-ph.GA},
       adsurl = {https://ui.adsabs.harvard.edu/abs/2014ApJ...784...83D}
}

@article{Casey12,
       author = {{Casey}, Caitlin M.},
        title = "{Far-infrared spectral energy distribution fitting for galaxies near and far}",
      journal = {Monthly Notices of the Royal Astronomical Society},
         year = 2012,
        month = {10},
       volume = {425},
       number = {4},
        pages = {3094-3103},
          doi = {10.1111/j.1365-2966.2012.21455.x},
archivePrefix = {arXiv},
       eprint = {1206.1595},
 primaryClass = {astro-ph.CO},
       adsurl = {https://ui.adsabs.harvard.edu/abs/2012MNRAS.425.3094C}
}

@inproceedings{DESI,
author = {Brenna Flaugher and Chris Bebek},
title = {{The Dark Energy Spectroscopic Instrument (DESI)}},
volume = {9147},
booktitle = {Ground-based and Airborne Instrumentation for Astronomy V},
editor = {Suzanne K. Ramsay and Ian S. McLean and Hideki Takami},
organization = {International Society for Optics and Photonics},
publisher = {SPIE},
pages = {91470S},
year = {2014},
doi = {10.1117/12.2057105},
URL = {https://doi.org/10.1117/12.2057105}
}

@misc{DESI_EDR,
  author       = {DESI Collaboration et al},
  title        = {{The Early Data Release of the Dark Energy Spectroscopic Instrument}},
  month        = {05},
  year         = 2023,
  publisher    = {Zenodo},
  version      = {v1.0},
  doi          = {10.5281/zenodo.7964162},
  url          = {https://doi.org/10.5281/zenodo.7964162}
}

@article{WISE,
       author = {{Wright}, Edward L. and {Eisenhardt}, Peter R.~M. and {Mainzer}, Amy K. and {Ressler}, Michael E. and {Cutri}, Roc M. and {Jarrett}, Thomas and {Kirkpatrick}, J. Davy and {Padgett}, Deborah and {McMillan}, Robert S. and {Skrutskie}, Michael and {Stanford}, S.~A. and {Cohen}, Martin and {Walker}, Russell G. and {Mather}, John C. and {Leisawitz}, David and {Gautier}, Thomas N., III and {McLean}, Ian and {Benford}, Dominic and {Lonsdale}, Carol J. and {Blain}, Andrew and {Mendez}, Bryan and {Irace}, William R. and {Duval}, Valerie and {Liu}, Fengchuan and {Royer}, Don and {Heinrichsen}, Ingolf and {Howard}, Joan and {Shannon}, Mark and {Kendall}, Martha and {Walsh}, Amy L. and {Larsen}, Mark and {Cardon}, Joel G. and {Schick}, Scott and {Schwalm}, Mark and {Abid}, Mohamed and {Fabinsky}, Beth and {Naes}, Larry and {Tsai}, Chao-Wei},
        title = "{The Wide-field Infrared Survey Explorer (WISE): Mission Description and Initial On-orbit Performance}",
      journal = {The Astronomical Journal},
         year = 2010,
        month = {12},
       volume = {140},
       number = {6},
        pages = {1868-1881},
          doi = {10.1088/0004-6256/140/6/1868},
archivePrefix = {arXiv},
       eprint = {1008.0031},
 primaryClass = {astro-ph.IM},
       adsurl = {https://ui.adsabs.harvard.edu/abs/2010AJ....140.1868W}
}

@article{LS_DR9_2019,
       author = {{Dey}, Arjun and {Schlegel}, David J. and {Lang}, Dustin and {Blum}, Robert and {Burleigh}, Kaylan and {Fan}, Xiaohui and {Findlay}, Joseph R. and {Finkbeiner}, Doug and {Herrera}, David and {Juneau}, St{\'e}phanie and {Landriau}, Martin and {Levi}, Michael and {McGreer}, Ian and {Meisner}, Aaron and {Myers}, Adam D. and {Moustakas}, John and {Nugent}, Peter and {Patej}, Anna and {Schlafly}, Edward F. and {Walker}, Alistair R. and {Valdes}, Francisco and {Weaver}, Benjamin A. and {Y{\`e}che}, Christophe and {Zou}, Hu and {Zhou}, Xu and {Abareshi}, Behzad and {Abbott}, T.~M.~C. and {Abolfathi}, Bela and {Aguilera}, C. and {Alam}, Shadab and {Allen}, Lori and {Alvarez}, A. and {Annis}, James and {Ansarinejad}, Behzad and {Aubert}, Marie and {Beechert}, Jacqueline and {Bell}, Eric F. and {BenZvi}, Segev Y. and {Beutler}, Florian and {Bielby}, Richard M. and {Bolton}, Adam S. and {Brice{\~n}o}, C{\'e}sar and {Buckley-Geer}, Elizabeth J. and {Butler}, Karen and {Calamida}, Annalisa and {Carlberg}, Raymond G. and {Carter}, Paul and {Casas}, Ricard and {Castander}, Francisco J. and {Choi}, Yumi and {Comparat}, Johan and {Cukanovaite}, Elena and {Delubac}, Timoth{\'e}e and {DeVries}, Kaitlin and {Dey}, Sharmila and {Dhungana}, Govinda and {Dickinson}, Mark and {Ding}, Zhejie and {Donaldson}, John B. and {Duan}, Yutong and {Duckworth}, Christopher J. and {Eftekharzadeh}, Sarah and {Eisenstein}, Daniel J. and {Etourneau}, Thomas and {Fagrelius}, Parker A. and {Farihi}, Jay and {Fitzpatrick}, Mike and {Font-Ribera}, Andreu and {Fulmer}, Leah and {G{\"a}nsicke}, Boris T. and {Gaztanaga}, Enrique and {George}, Koshy and {Gerdes}, David W. and {Gontcho}, Satya Gontcho A. and {Gorgoni}, Claudio and {Green}, Gregory and {Guy}, Julien and {Harmer}, Diane and {Hernandez}, M. and {Honscheid}, Klaus and {Huang}, Lijuan Wendy and {James}, David J. and {Jannuzi}, Buell T. and {Jiang}, Linhua and {Joyce}, Richard and {Karcher}, Armin and {Karkar}, Sonia and {Kehoe}, Robert and {Kneib}, Jean-Paul and {Kueter-Young}, Andrea and {Lan}, Ting-Wen and {Lauer}, Tod R. and {Le Guillou}, Laurent and {Le Van Suu}, Auguste and {Lee}, Jae Hyeon and {Lesser}, Michael and {Perreault Levasseur}, Laurence and {Li}, Ting S. and {Mann}, Justin L. and {Marshall}, Robert and {Mart{\'\i}nez-V{\'a}zquez}, C.~E. and {Martini}, Paul and {du Mas des Bourboux}, H{\'e}lion and {McManus}, Sean and {Meier}, Tobias Gabriel and {M{\'e}nard}, Brice and {Metcalfe}, Nigel and {Mu{\~n}oz-Guti{\'e}rrez}, Andrea and {Najita}, Joan and {Napier}, Kevin and {Narayan}, Gautham and {Newman}, Jeffrey A. and {Nie}, Jundan and {Nord}, Brian and {Norman}, Dara J. and {Olsen}, Knut A.~G. and {Paat}, Anthony and {Palanque-Delabrouille}, Nathalie and {Peng}, Xiyan and {Poppett}, Claire L. and {Poremba}, Megan R. and {Prakash}, Abhishek and {Rabinowitz}, David and {Raichoor}, Anand and {Rezaie}, Mehdi and {Robertson}, A.~N. and {Roe}, Natalie A. and {Ross}, Ashley J. and {Ross}, Nicholas P. and {Rudnick}, Gregory and {Safonova}, Sasha and {Saha}, Abhijit and {S{\'a}nchez}, F. Javier and {Savary}, Elodie and {Schweiker}, Heidi and {Scott}, Adam and {Seo}, Hee-Jong and {Shan}, Huanyuan and {Silva}, David R. and {Slepian}, Zachary and {Soto}, Christian and {Sprayberry}, David and {Staten}, Ryan and {Stillman}, Coley M. and {Stupak}, Robert J. and {Summers}, David L. and {Sien Tie}, Suk and {Tirado}, H. and {Vargas-Maga{\~n}a}, Mariana and {Vivas}, A. Katherina and {Wechsler}, Risa H. and {Williams}, Doug and {Yang}, Jinyi and {Yang}, Qian and {Yapici}, Tolga and {Zaritsky}, Dennis and {Zenteno}, A. and {Zhang}, Kai and {Zhang}, Tianmeng and {Zhou}, Rongpu and {Zhou}, Zhimin},
        title = "{Overview of the DESI Legacy Imaging Surveys}",
      journal = {The Astronomical Journal},
         year = 2019,
        month = {05},
       volume = {157},
       number = {5},
          eid = {168},
        pages = {168},
          doi = {10.3847/1538-3881/ab089d},
archivePrefix = {arXiv},
       eprint = {1804.08657},
 primaryClass = {astro-ph.IM},
       adsurl = {https://ui.adsabs.harvard.edu/abs/2019AJ....157..168D}
}

@article{BASS,
       author = {{Zou}, Hu and {Zhou}, Xu and {Fan}, Xiaohui and {Zhang}, Tianmeng and {Zhou}, Zhimin and {Nie}, Jundan and {Peng}, Xiyan and {McGreer}, Ian and {Jiang}, Linhua and {Dey}, Arjun and {Fan}, Dongwei and {He}, Boliang and {Jiang}, Zhaoji and {Lang}, Dustin and {Lesser}, Michael and {Ma}, Jun and {Mao}, Shude and {Schlegel}, David and {Wang}, Jiali},
        title = "{Project Overview of the Beijing-Arizona Sky Survey}",
      journal = {Publications of the Astronomical Society of the Pacific},
         year = 2017,
        month = {06},
       volume = {129},
       number = {976},
        pages = {064101},
          doi = {10.1088/1538-3873/aa65ba},
archivePrefix = {arXiv},
       eprint = {1702.03653},
 primaryClass = {astro-ph.GA},
       adsurl = {https://ui.adsabs.harvard.edu/abs/2017PASP..129f4101Z}
}

@article{SDSS,
       author = {{Almeida}, Andr{\'e}s and {Anderson}, Scott F. and {Argudo-Fern{\'a}ndez}, Maria and {Badenes}, Carles and {Barger}, Kat and {Barrera-Ballesteros}, Jorge K. and {Bender}, Chad F. and {Benitez}, Erika and {Besser}, Felipe and {Bird}, Jonathan C. and {Bizyaev}, Dmitry and {Blanton}, Michael R. and {Bochanski}, John and {Bovy}, Jo and {Brandt}, William Nielsen and {Brownstein}, Joel R. and {Buchner}, Johannes and {Bulbul}, Esra and {Burchett}, Joseph N. and {Cano D{\'\i}az}, Mariana and {Carlberg}, Joleen K. and {Casey}, Andrew R. and {Chandra}, Vedant and {Cherinka}, Brian and {Chiappini}, Cristina and {Coker}, Abigail A. and {Comparat}, Johan and {Conroy}, Charlie and {Contardo}, Gabriella and {Cortes}, Arlin and {Covey}, Kevin and {Crane}, Jeffrey D. and {Cunha}, Katia and {Dabbieri}, Collin and {Davidson}, James W. and {Davis}, Megan C. and {de Andrade Queiroz}, Anna Barbara and {De Lee}, Nathan and {M{\'e}ndez Delgado}, Jos{\'e} Eduardo and {Demasi}, Sebastian and {Di Mille}, Francesco and {Donor}, John and {Dow}, Peter and {Dwelly}, Tom and {Eracleous}, Mike and {Eriksen}, Jamey and {Fan}, Xiaohui and {Farr}, Emily and {Frederick}, Sara and {Fries}, Logan and {Frinchaboy}, Peter and {G{\"a}nsicke}, Boris T. and {Ge}, Junqiang and {Gonz{\'a}lez {\'A}vila}, Consuelo and {Grabowski}, Katie and {Grier}, Catherine and {Guiglion}, Guillaume and {Gupta}, Pramod and {Hall}, Patrick and {Hawkins}, Keith and {Hayes}, Christian R. and {Hermes}, J.~J. and {Hern{\'a}ndez-Garc{\'\i}a}, Lorena and {Hogg}, David W. and {Holtzman}, Jon A. and {Ibarra-Medel}, Hector Javier and {Ji}, Alexander and {Jofre}, Paula and {Johnson}, Jennifer A. and {Jones}, Amy M. and {Kinemuchi}, Karen and {Kluge}, Matthias and {Koekemoer}, Anton and {Kollmeier}, Juna A. and {Kounkel}, Marina and {Krishnarao}, Dhanesh and {Krumpe}, Mirko and {Lacerna}, Ivan and {Lago}, Paulo Jakson Assuncao and {Laporte}, Chervin and {Liu}, Chao and {Liu}, Ang and {Liu}, Xin and {Lopes}, Alexandre Roman and {Macktoobian}, Matin and {Majewski}, Steven R. and {Malanushenko}, Viktor and {Maoz}, Dan and {Masseron}, Thomas and {Masters}, Karen L. and {Matijevic}, Gal and {McBride}, Aidan and {Medan}, Ilija and {Merloni}, Andrea and {Morrison}, Sean and {Myers}, Natalie and {M{\'e}sz{\'a}ros}, Szabolcs and {Negrete}, C. Alenka and {Nidever}, David L. and {Nitschelm}, Christian and {Oravetz}, Daniel and {Oravetz}, Audrey and {Pan}, Kaike and {Peng}, Yingjie and {Pinsonneault}, Marc H. and {Pogge}, Rick and {Qiu}, Dan and {Ramirez}, Solange V. and {Rix}, Hans-Walter and {Fern{\'a}ndez Rosso}, Daniela and {Runnoe}, Jessie and {Salvato}, Mara and {Sanchez}, Sebastian F. and {Santana}, Felipe A. and {Saydjari}, Andrew and {Sayres}, Conor and {Schlaufman}, Kevin C. and {Schneider}, Donald P. and {Schwope}, Axel and {Serna}, Javier and {Shen}, Yue and {Sobeck}, Jennifer and {Song}, Ying-Yi and {Souto}, Diogo and {Spoo}, Taylor and {Stassun}, Keivan G. and {Steinmetz}, Matthias and {Straumit}, Ilya and {Stringfellow}, Guy and {S{\'a}nchez-Gallego}, Jos{\'e} and {Taghizadeh-Popp}, Manuchehr and {Tayar}, Jamie and {Thakar}, Ani and {Tissera}, Patricia B. and {Tkachenko}, Andrew and {Hernandez Toledo}, Hector and {Trakhtenbrot}, Benny and {Fern{\'a}ndez-Trincado}, Jos{\'e} G. and {Troup}, Nicholas and {Trump}, Jonathan R. and {Tuttle}, Sarah and {Ulloa}, Natalie and {Vazquez-Mata}, Jose Antonio and {Vera Alfaro}, Pablo and {Villanova}, Sandro and {Wachter}, Stefanie and {Weijmans}, Anne-Marie and {Wheeler}, Adam and {Wilson}, John and {Wojno}, Leigh and {Wolf}, Julien and {Xue}, Xiang-Xiang and {Ybarra}, Jason E. and {Zari}, Eleonora and {Zasowski}, Gail},
        title = "{The Eighteenth Data Release of the Sloan Digital Sky Surveys: Targeting and First Spectra from SDSS-V}",
      journal = {The Astrophysical Journal Supplement Series},
         year = 2023,
        month = {08},
       volume = {267},
       number = {2},
          eid = {44},
        pages = {44},
          doi = {10.3847/1538-4365/acda98},
archivePrefix = {arXiv},
       eprint = {2301.07688},
 primaryClass = {astro-ph.GA},
       adsurl = {https://ui.adsabs.harvard.edu/abs/2023ApJS..267...44A}
}

@article{GAMA,
       author = {{Driver}, Simon P. and {Bellstedt}, Sabine and {Robotham}, Aaron S.~G. and {Baldry}, Ivan K. and {Davies}, Luke J. and {Liske}, Jochen and {Obreschkow}, Danail and {Taylor}, Edward N. and {Wright}, Angus H. and {Alpaslan}, Mehmet and {Bamford}, Steven P. and {Bauer}, Amanda E. and {Bland-Hawthorn}, Joss and {Bilicki}, Maciej and {Bravo}, Mat{\'\i}as and {Brough}, Sarah and {Casura}, Sarah and {Cluver}, Michelle E. and {Colless}, Matthew and {Conselice}, Christopher J. and {Croom}, Scott M. and {de Jong}, Jelte and {D'Eugenio}, Franceso and {De Propris}, Roberto and {Dogruel}, Burak and {Drinkwater}, Michael J. and {Dvornik}, Andrej and {Farrow}, Daniel J. and {Frenk}, Carlos S. and {Giblin}, Benjamin and {Graham}, Alister W. and {Grootes}, Meiert W. and {Gunawardhana}, Madusha L.~P. and {Hashemizadeh}, Abdolhosein and {H{\"a}u{\ss}ler}, Boris and {Heymans}, Catherine and {Hildebrandt}, Hendrik and {Holwerda}, Benne W. and {Hopkins}, Andrew M. and {Jarrett}, Tom H. and {Heath Jones}, D. and {Kelvin}, Lee S. and {Koushan}, Soheil and {Kuijken}, Konrad and {Lara-L{\'o}pez}, Maritza A. and {Lange}, Rebecca and {L{\'o}pez-S{\'a}nchez}, {\'A}ngel R. and {Loveday}, Jon and {Mahajan}, Smriti and {Meyer}, Martin and {Moffett}, Amanda J. and {Napolitano}, Nicola R. and {Norberg}, Peder and {Owers}, Matt S. and {Radovich}, Mario and {Raouf}, Mojtaba and {Peacock}, John A. and {Phillipps}, Steven and {Pimbblet}, Kevin A. and {Popescu}, Cristina and {Said}, Khaled and {Sansom}, Anne E. and {Seibert}, Mark and {Sutherland}, Will J. and {Thorne}, Jessica E. and {Tuffs}, Richard J. and {Turner}, Ryan and {van der Wel}, Arjen and {van Kampen}, Eelco and {Wilkins}, Steve M.},
        title = "{Galaxy And Mass Assembly (GAMA): Data Release 4 and the z < 0.1 total and z < 0.08 morphological galaxy stellar mass functions}",
      journal = {Monthly Notices of the Royal Astronomical Society},
         year = 2022,
        month = {06},
       volume = {513},
       number = {1},
        pages = {439-467},
          doi = {10.1093/mnras/stac472},
archivePrefix = {arXiv},
       eprint = {2203.08539},
 primaryClass = {astro-ph.GA},
       adsurl = {https://ui.adsabs.harvard.edu/abs/2022MNRAS.513..439D}
}

@article{Salim16,
       author = {{Salim}, Samir and {Lee}, Janice C. and {Janowiecki}, Steven and {da Cunha}, Elisabete and {Dickinson}, Mark and {Boquien}, M{\'e}d{\'e}ric and {Burgarella}, Denis and {Salzer}, John J. and {Charlot}, St{\'e}phane},
        title = "{GALEX-SDSS-WISE Legacy Catalog (GSWLC): Star Formation Rates, Stellar Masses, and Dust Attenuations of 700,000 Low-redshift Galaxies}",
      journal = {The Astrophysical Journal Supplement Series},
         year = 2016,
        month = {11},
       volume = {227},
       number = {1},
          eid = {2},
        pages = {2},
          doi = {10.3847/0067-0049/227/1/2},
archivePrefix = {arXiv},
       eprint = {1610.00712},
 primaryClass = {astro-ph.GA},
       adsurl = {https://ui.adsabs.harvard.edu/abs/2016ApJS..227....2S}
}

@article{Salim18,
       author = {{Salim}, Samir and {Boquien}, M{\'e}d{\'e}ric and {Lee}, Janice C.},
        title = "{Dust Attenuation Curves in the Local Universe: Demographics and New Laws for Star-forming Galaxies and High-redshift Analogs}",
      journal = {The Astrophysical Journal},
         year = 2018,
        month = {05},
       volume = {859},
       number = {1},
          eid = {11},
        pages = {11},
          doi = {10.3847/1538-4357/aabf3c},
archivePrefix = {arXiv},
       eprint = {1804.05850},
 primaryClass = {astro-ph.GA},
       adsurl = {https://ui.adsabs.harvard.edu/abs/2018ApJ...859...11S}
}

@article{MESA,
doi = {10.1088/0067-0049/192/1/3},
url = {https://dx.doi.org/10.1088/0067-0049/192/1/3},
year = {2010},
month = {12},
publisher = {The American Astronomical Society},
volume = {192},
number = {1},
pages = {3},
author = {Bill Paxton and Lars Bildsten and Aaron Dotter and Falk Herwig and Pierre Lesaffre and Frank Timmes},
title = {MODULES FOR EXPERIMENTS IN STELLAR ASTROPHYSICS (MESA)},
journal = {The Astrophysical Journal Supplement Series}
}

@article{MIST,
doi = {10.3847/0004-637X/823/2/102},
url = {https://dx.doi.org/10.3847/0004-637X/823/2/102},
year = {2016},
month = {05},
publisher = {The American Astronomical Society},
volume = {823},
number = {2},
pages = {102},
author = {Jieun Choi and Aaron Dotter and Charlie Conroy and Matteo Cantiello and Bill Paxton and Benjamin D. Johnson},
title = {MESA ISOCHRONES AND STELLAR TRACKS (MIST). I. SOLAR-SCALED MODELS},
journal = {The Astrophysical Journal}
}

@article{MILES,
    author = {Sánchez-Blázquez, P. and Peletier, R. F. and Jiménez-Vicente, J. and Cardiel, N. and Cenarro, A. J. and Falcón-Barroso, J. and Gorgas, J. and Selam, S. and Vazdekis, A.},
    title = "{Medium-resolution Isaac Newton Telescope library of empirical spectra}",
    journal = {Monthly Notices of the Royal Astronomical Society},
    volume = {371},
    number = {2},
    pages = {703-718},
    year = {2006},
    month = {08},
    issn = {0035-8711},
    doi = {10.1111/j.1365-2966.2006.10699.x},
    url = {https://doi.org/10.1111/j.1365-2966.2006.10699.x},
    eprint = {https://academic.oup.com/mnras/article-pdf/371/2/703/3443797/mnras0371-0703.pdf},
}

@article{MILES_up,
       author = {{Falc{\'o}n-Barroso}, J. and {S{\'a}nchez-Bl{\'a}zquez}, P. and {Vazdekis}, A. and {Ricciardelli}, E. and {Cardiel}, N. and {Cenarro}, A.~J. and {Gorgas}, J. and {Peletier}, R.~F.},
        title = "{An updated MILES stellar library and stellar population models}",
      journal = {Astronomy and Astrophysics},
         year = 2011,
        month = {08},
       volume = {532},
          eid = {A95},
        pages = {A95},
          doi = {10.1051/0004-6361/201116842},
archivePrefix = {arXiv},
       eprint = {1107.2303},
 primaryClass = {astro-ph.CO},
       adsurl = {https://ui.adsabs.harvard.edu/abs/2011AandA...532A..95F}
}

@article{Calzetti00,
       author = {{Calzetti}, Daniela and {Armus}, Lee and {Bohlin}, Ralph C. and {Kinney}, Anne L. and {Koornneef}, Jan and {Storchi-Bergmann}, Thaisa},
        title = "{The Dust Content and Opacity of Actively Star-forming Galaxies}",
      journal = {The Astrophysical Journal},
         year = 2000,
        month = {04},
       volume = {533},
       number = {2},
        pages = {682-695},
          doi = {10.1086/308692},
archivePrefix = {arXiv},
       eprint = {astro-ph/9911459},
 primaryClass = {astro-ph},
       adsurl = {https://ui.adsabs.harvard.edu/abs/2000ApJ...533..682C}
}

@article{Wild11,
       author = {{Wild}, Vivienne and {Charlot}, St{\'e}phane and {Brinchmann}, Jarle and {Heckman}, Timothy and {Vince}, Oliver and {Pacifici}, Camilla and {Chevallard}, Jacopo},
        title = "{Empirical determination of the shape of dust attenuation curves in star-forming galaxies}",
      journal = {Monthly Notices of the Royal Astronomical Society},
         year = 2011,
        month = {11},
       volume = {417},
       number = {3},
        pages = {1760-1786},
          doi = {10.1111/j.1365-2966.2011.19367.x},
archivePrefix = {arXiv},
       eprint = {1106.1646},
 primaryClass = {astro-ph.CO},
       adsurl = {https://ui.adsabs.harvard.edu/abs/2011MNRAS.417.1760W}
}

@article{Reddy15,
       author = {{Reddy}, Naveen A. and {Kriek}, Mariska and {Shapley}, Alice E. and {Freeman}, William R. and {Siana}, Brian and {Coil}, Alison L. and {Mobasher}, Bahram and {Price}, Sedona H. and {Sanders}, Ryan L. and {Shivaei}, Irene},
        title = "{The MOSDEF Survey: Measurements of Balmer Decrements and the Dust Attenuation Curve at Redshifts z \raisebox{-0.5ex}\textasciitilde 1.4-2.6}",
      journal = {The Astrophysical Journal},
         year = 2015,
        month = {06},
       volume = {806},
       number = {2},
          eid = {259},
        pages = {259},
          doi = {10.1088/0004-637X/806/2/259},
archivePrefix = {arXiv},
       eprint = {1504.02782},
 primaryClass = {astro-ph.GA},
       adsurl = {https://ui.adsabs.harvard.edu/abs/2015ApJ...806..259R}
}

@article{Reddy16,
       author = {{Reddy}, Naveen A. and {Steidel}, Charles C. and {Pettini}, Max and {Bogosavljevi{\'c}}, Milan},
        title = "{Spectroscopic Measurements of the Far-Ultraviolet Dust Attenuation Curve at z {\ensuremath{\sim}} 3}",
      journal = {The Astrophysical Journal},
         year = 2016,
        month = {09},
       volume = {828},
       number = {2},
          eid = {107},
        pages = {107},
          doi = {10.3847/0004-637X/828/2/107},
archivePrefix = {arXiv},
       eprint = {1606.00434},
 primaryClass = {astro-ph.GA},
       adsurl = {https://ui.adsabs.harvard.edu/abs/2016ApJ...828..107R}
}

@article{LoFaro17,
       author = {{Lo Faro}, B. and {Buat}, V. and {Roehlly}, Y. and {Alvarez-Marquez}, J. and {Burgarella}, D. and {Silva}, L. and {Efstathiou}, A.},
        title = "{Characterizing the UV-to-NIR shape of the dust attenuation curve of IR luminous galaxies up to z {\ensuremath{\sim}} 2}",
      journal = {Monthly Notices of the Royal Astronomical Society},
         year = 2017,
        month = {12},
       volume = {472},
       number = {2},
        pages = {1372-1391},
          doi = {10.1093/mnras/stx1901},
archivePrefix = {arXiv},
       eprint = {1707.09805},
 primaryClass = {astro-ph.GA},
       adsurl = {https://ui.adsabs.harvard.edu/abs/2017MNRAS.472.1372L}
}

@article{THEMIS,
       author = {{Jones}, A.~P. and {K{\"o}hler}, M. and {Ysard}, N. and {Bocchio}, M. and {Verstraete}, L.},
        title = "{The global dust modelling framework THEMIS}",
      journal = {Astronomy and Astrophysics},
         year = 2017,
        month = {06},
       volume = {602},
          eid = {A46},
        pages = {A46},
          doi = {10.1051/0004-6361/201630225},
archivePrefix = {arXiv},
       eprint = {1703.00775},
 primaryClass = {astro-ph.GA},
       adsurl = {https://ui.adsabs.harvard.edu/abs/2017AandA...602A..46J}
}

@article{GALEX,
doi = {10.1086/520512},
url = {https://dx.doi.org/10.1086/520512},
year = {2007},
month = {12},
publisher = {The American Astronomical Society},
volume = {173},
number = {2},
pages = {682},
author = {Patrick Morrissey and Tim Conrow and Tom A. Barlow and Todd Small and Mark Seibert and Ted K. Wyder and Tamás Budavári and Stephane Arnouts and Peter G. Friedman and Karl Forster and D. Christopher Martin and Susan G. Neff and David Schiminovich and Luciana Bianchi and José Donas and Timothy M. Heckman and Young-Wook Lee and Barry F. Madore and Bruno Milliard and R. Michael Rich and Alex S. Szalay and Barry Y. Welsh and Sukyoung K. Yi},
title = {The Calibration and Data Products of GALEX},
journal = {The Astrophysical Journal Supplement Series}
}

@article{Gal_MS,
    author = {Popesso, P and Concas, A and Cresci, G and Belli, S and Rodighiero, G and Inami, H and Dickinson, M and Ilbert, O and Pannella, M and Elbaz, D},
    title = "{The main sequence of star-forming galaxies across cosmic times}",
    journal = {Monthly Notices of the Royal Astronomical Society},
    volume = {519},
    number = {1},
    pages = {1526-1544},
    year = {2022},
    month = {11},
    issn = {0035-8711},
    doi = {10.1093/mnras/stac3214},
    url = {https://doi.org/10.1093/mnras/stac3214},
    eprint = {https://academic.oup.com/mnras/article-pdf/519/1/1526/48447685/stac3214.pdf},
}

@article{LambdarPhotom,
    author = {Wright, A. H. and Robotham, A. S. G. and Bourne, N. and Driver, S. P. and Dunne, L. and Maddox, S. J. and Alpaslan, M. and Andrews, S. K. and Bauer, A. E. and Bland-Hawthorn, J. and Brough, S. and Brown, M. J. I. and Clarke, C. and Cluver, M. and Davies, L. J. M. and Grootes, M. W. and Holwerda, B. W. and Hopkins, A. M. and Jarrett, T. H. and Kafle, P. R. and Lange, R. and Liske, J. and Loveday, J. and Moffett, A. J. and Norberg, P. and Popescu, C. C. and Smith, M. and Taylor, E. N. and Tuffs, R. J. and Wang, L. and Wilkins, S. M.},
    title = "{Galaxy And Mass Assembly: accurate panchromatic photometry from optical priors using lambdar}",
    journal = {Monthly Notices of the Royal Astronomical Society},
    volume = {460},
    number = {1},
    pages = {765-801},
    year = {2016},
    month = {04},
    issn = {0035-8711},
    doi = {10.1093/mnras/stw832},
    url = {https://doi.org/10.1093/mnras/stw832},
    eprint = {https://academic.oup.com/mnras/article-pdf/460/1/765/8114174/stw832.pdf},
}
